\documentclass[aps,showpacs,onecolumn,floatfix,amsmath,amssymb,nofootinbib,prd]{revtex4-1}
\usepackage{amssymb}
\usepackage{amsmath}
\usepackage{epsfig}
\usepackage{graphicx}
\usepackage{latexsym}
\usepackage{bm,latexsym}
\usepackage{mathrsfs}

\def\Journal#1#2#3#4{{#1} {\bf #2}, #3 (#4)}

\def\AJ{{Astron. J.}}

\def\APJ{{Astrophys. J.}}
\def\APJS{{Astrophys. J. Suppl. Ser.}}
\def\ARAA{{Annu. Rev. Astron. Astrophys.}}

\def\CQG{{Class. Quant. Grav.}}
\def\EPJC{{Eur. Phys. J. C}}
\def\GRG{{Gen. Relativ. Gravit.}}
\def\IJMPD{{Int. Jour. Mod. Phys. D}}
\def\JCAP{{JCAP}}

\def\FCP{{Fund. CosmicPhys.}}

\def\LRR{{Living Rev. Rel.}} 
\def\MNRAS{{Mon. Not. Roy. Astron. Soc.}} 
\def\NAT{{Nature}}

\def\NPBPS{Nucl. Phys. B (Proc. Suppl.}
\def\PLB{{Phys. Lett.}  B}

\def\PRL{Phys. Rev. Lett.}

\def\PRD{{Phys. Rev.} D}
\def\PR{{Phys. Rep.}}
\def\PTP{{Prog. Theor. Phys.}}
\def\SCI{{Science}}

\def\RMP{{Rev. Mod. Phys.}}

\begin{document}


\title{$f(R)$ cosmology revisited}

\author{Luisa G. Jaime$^{1,2}$}
\email{luisa@nucleares.unam.mx}

\author{Leonardo Pati\~no$^2$}
\email{leopj@ciencias.unam.mx}

\author{Marcelo Salgado$^1$}
\email{marcelo@nucleares.unam.mx}

\affiliation{$^1$Instituto de Ciencias Nucleares, Universidad Nacional
Aut\'onoma de M\'exico, A.P. 70-543, M\'exico D.F. 04510, M\'exico \\
$^2$ Facultad de Ciencias, Universidad Nacional
Aut\'onoma de M\'exico, A.P. 50-542, M\'exico D.F. 04510, M\'exico }


\date{\today}


\begin{abstract}
We consider a class of metric $f(R)$ modified gravity theories, analyze them in the context of a Friedmann--Robertson--Walker 
cosmology and confront the results with some of the known constraints imposed by observations. In particular, we focus in correctly 
reproducing the matter and effective cosmological constant eras, the age of the Universe, 
and supernovae data. Our analysis differs in many respects from previous studies. First, we avoid any transformation to a scalar-tensor theory in order 
to be exempted of any potential pathologies (e.g. multivalued scalar potentials) and also to evade any unnecessary discussion regarding frames 
(i.e. Einstein {\it .vs.} Jordan). Second, based on a robust approach, we recast the cosmology equations as an initial 
value problem subject to a modified Hamiltonian constraint. Third, we solve the equations numerically where the Ricci scalar itself is one of the 
variables, and use the constraint equation to monitor the accuracy of the solutions. We compute the ``equation of state'' (EOS) associated with the 
modifications of gravity using several inequivalent definitions that have been proposed in the past and analyze it in detail. We argue that 
one of these definitions has the best features. 
In particular, we present the EOS around the so called ``phantom divide'' boundary and compare it with previous findings.
\bigskip

\end{abstract}




\maketitle


\section{Introduction}
\label{sec:introduction}

Astronomical observations based on type Ia supernovae (SNIa) together with the assumption that the Universe is homogeneous and isotropic 
at large scales led to the conclusion that the Universe is currently expanding in an accelerated way~\cite{Perlmutter1999,Riess1998,Amanullah2010}. 
This phenomenon can be most easily explained by appealing to the existence of a 
cosmological constant  $\Lambda$ (sometimes termed {\it dark energy}). This constant along with the introduction of dark matter (DM) apparently needed in many 
regions of the Universe (galaxies and clusters) have originated what is called today the $\Lambda CDM$ paradigm. This paradigm has also successfully explained most of 
the current details of the Cosmic Background Radiation (CBR or CMB) in the framework of general relativity~\cite{WMAP}, as well as other 
important features of the Universe at large scales~\cite{LSS} (for a thorough review see Ref.~\cite{Weinberg2012}).

However, despite of the simplicity and success of this paradigm, several theoretical as well as epistemological arguments have been put forward 
as objections against such a simple model of the Universe. For instance, as concerns the DM hypothesis, one of the the main criticisms 
is that its nature (i.e. its quantum and classical properties) is not well understood (if at all) yet. That is, apart from the gravitational 
evidence, there is no further strong reason supporting its existence. Since several experiments have failed so far to detect the proposed DM particles, 
skepticism keeps growing in this direction. 
On the other hand, the cosmological constant has been historically regarded as ``suspicious''  
by several detractors (including Einstein himself; see Refs.~\cite{Lambda} for a review), although some of its apparent drawbacks are based more on 
prejudices than on strong and well grounded physical arguments~\cite{Bianchi2010}. 
In any case, the discomfort that $\Lambda$ has produced in the spirit of some people has led to consider more complicated alternatives, of which, 
is fair to say, none is regarded today as a more serious candidate for dark energy (DE) than $\Lambda$ (the BigBOSS experiment~\cite{BB} has been 
designed to shed light in this direction). Among these alternatives are the so called modified 
theories of gravity (MTG) as opposed to general relativity (GR). Some of these theories have been also proposed to substitute DM and even as models for 
inflation. Perhaps the most popular MTG over the past ten years and the one we focus in this article are $f(R)$ metric theories, where an 
{\it a priori} arbitrary function of the Ricci scalar $R$ replaces $R$ itself in the gravitational Lagrangian. 

This kind of MTG were conceived originally in order to create a late accelerated effect without a cosmological constant or as inflationary model 
without an extra scalar field (see Refs.~\cite{f(R),Capozziello2008a,Sotiriou2010,deFelice2010} for a detailed review). Notwithstanding, despite of the promising 
$f(R)$ models first proposed to replace $\Lambda$ \cite{accexp}, a cumulative evidence, both theoretical and observational, has been found against 
most of them. However, new models have been proposed to overcome the initial difficulties, some better motivated that others but none introducing a new fundamental 
principle that can be used as a guiding line; indeed they have rather been constructed by trial and error. 
The simplest (non-trivial) choice $f(R)=R$ was historically favored by Einstein since mathematically led to second order partial differential 
equations (PDE's) which could easily reduce to the Newtonian theory in the week field limit.

General mathematical and physical conditions are usually demanded in order to avoid pathologies in the models. For instance, 
the conditions $f_{RR}>0$ and $f_R>0$ (where the subindex indicate derivative with respect to $R$) seem to be required for stability considerations 
and to ensure a positive definite effective gravitational constant, respectively. It is however not clear if those conditions are really necessary, and 
in many studies they are not imposed. Therefore, in most cases, ``handcraft'' has been used to design a particular $f(R)$ model 
based on heuristic arguments that might account for the actual phenomenology when the full fledge model is submitted to a detailed scrutiny. 
The general trend so far is that no single model is able to explain most of the current observations, but only 
some aspects of them. That is, most of the $f(R)$ models fail miserably when they are preempted as models for all the dark substance (both DE and DM) and when 
taking into account the Solar System tests as well. Even when considered only as DE models, most of them fail, with the exception of some notable 
cases. Of course it could well happen (but perhaps not very desirable) that dark matter, dark energy and modifications of the 
laws of gravity in some combination are required by nature in order to fully understand our Universe. In fact, this is the approach we 
pursue here at the cosmological level except that we do not include explicitly a cosmological constant, but rather, the models themselves 
give rise to an effective $\Lambda$ which vary slightly around the required value at late (i.e. present) epochs of the 
Universe. We then consider $f(R)$ theories simply as a model for explaining the accelerated expansion of the Universe and 
introduce a dark matter component in the same way as in the $\Lambda CDM$ paradigm. Nevertheless, things turn out to be not so simple, given that the 
proposed models can disturb the successes of GR. Any proposed specific $f(R)$ model has not only to satisfy the cosmological observations, 
but all the gravitational observations at all scales. These include the Solar System experiments, the existence of physically acceptable compact 
objects, the binary pulsar, etc. As today, there is no single MTG model that replaces successfully GR and explains as should be, 
all the observations for which it was designed originally.

After the first  $f(R)$ models were proposed to explain the accelerated expansion of the Universe, among them 
the ``historic'' $f(R)= R - \mu^4/R$, a sequence of papers appeared where the constraints imposed by the Solar System 
were taken into account \cite{solarsystem1}. Without reaching a clear consensus on the issue, it seemed that such models were not viable. 
One of the arguments put forward to establish that conclusion 
was based on the fact that such theories can be shown to be dynamically equivalent to a Brans-Dicke (BD) theory with $\omega=0$. Since such 
a value for $\omega$ in these theories gives rise to a post-Newtonian parameter $\gamma=1/2$, which conflict with $\gamma\sim 1$ favored by the Solar 
System tests, then at first sight the analysis suggested that all $f(R)$ theories were excluded blatantly as viable theories. Much later, 
it was recognized that such an argument should be used with care in view that $f(R)$ theories are not equivalent to the standard BD theory 
with $\omega=0$, but to a BD theory with a potential. Therefore, depending on the mass of the effective scalar, the theory at hand could 
pass or fail the Solar System tests~\cite{solarsystem2}. Although it is now recognized that many of the $f(R)$ models give rise to $\gamma\sim 1/2$, and are 
therefore ruled out, some others, due to the effective mass of the scalar, might produce a successful phenomenology. 
This success depends on whether or not the scalar field which is associated with the model at hand can act as a {\it chameleon}~
\cite{chameleon,Hu2007}, a mechanism that appears in some scalar-tensor theories of gravity which allows them to 
satisfy the local tests and the possibility of producing the required cosmological effects~\cite{Khoury}.
 
Now, as concerns the cosmological restrictions on these theories, Amendola {\it et al.}~\cite{Amendola2007a,Amendola2007b} have devised criteria of quite general 
applicability that allows to discard many of the proposed $f(R)$ models. In short, their analysis shows for a large class of models that they either produce 
an accelerated expansion at recent times but fail to generate a correct matter-dominated era 
(the scale factor behaves as radiation in GR) or the opposite. In fact, when viewed from the past to the present, many of such models cross from
the radiation-dominated era (deceleration epoch) to an effective dark-energy era (acceleration epoch) with a very short or not even existent
matter-domination era. Such models are therefore incompatible with the CBR observations, the age of the Universe and the structure 
formation at large scales~\cite{Amendola2007a,Amendola2007b}. 
This issue was, however, not free of debate either~\cite{Capozziello2007,Amendola2007c}.

To make things even more confusing in this matter, additional skepticism was raised about the viability of such kind of theories when a further test was performed 
on several cosmologically successful $f(R)$ models. This time, the 
test consisted in analyzing the possibility that such models allowed the construction of solutions representing realistic (or at least idealized) neutron stars. 
In a first attempt to do so, Kobayashi \& Maeda~\cite{Kobayashi2008} showed that idealized neutron stars 
(specifically, incompressible compact objects) were not allowed by the Starobinsky model~\cite{Starobinsky2007} 
since a singularity in the Ricci scalar developed within the 
object. Later, Babichev \& Langlois~\cite{Babichev2009} criticized such conclusion and argued that the singularity was only due to the use of an incompressible 
fluid and when a similar situation was analyzed with a compressible gas (e.g. a polytrope) no singularity was found. Finally, 
Upadhye \& Hu~\cite{Upadhye2009} argued that a chameleon effect was the responsible of avoiding the formation of singularities in compact objects and not 
the use of more realistic equation of state.

In a more recent work by us~\cite{Jaime2011}, while we arrived to the same primary conclusion of Refs.~\cite{Babichev2009,Upadhye2009} 
on that no singularities were necessarily formed, 
we criticized some very basic aspects of the analysis common to all of the three above investigations concerning the existence of neutron 
stars~\cite{Kobayashi2008,Babichev2009,Upadhye2009}. To be specific here let us mention that their analysis relied fundamentally on the fact 
that a potential associated with a scalar-tensor counterpart of the Starobinsky model could drive the scalar field to a point where the corresponding 
Ricci scalar diverged. The main point of our criticism, noticed earlier in~\cite{Goheer2009,Miranda2009}, 
is that such a potential has unpleasant features as it is multivalued and therefore, 
the transformation to the STT is not well defined. Since the conclusions reached on those papers depend crucially on the use of such a pathological 
potential we consider them not very trustworthy, even if the dynamics is supposed to take place in a region where the potential is single valued. 
Moreover, due to the fact that many of the controversies regarding this subject have been the result of using the STT in the Einstein or Jordan frames 
(see also Ref.~\cite{Multamaki2006} for a similar criticism), we strongly suggested to abandon such an approach and treat $f(R)$ models in all applications 
without performing the transformation to any frame of STT. In the particular 
case of static and spherically symmetric spacetimes, we devised a very transparent and simple approach that allowed us to deal with compact objects. 
Using a particular case of the Starobinsky model we concluded that no singularities were found. We also studied the Miranda 
{\it et al.} model~\cite{Miranda2009} (which was previously shown to be free of singularities following the STT approach but with a single-valued 
potential), and arrived to the same conclusion. We emphasized the advantages of using that robust approach over the STT technique, and we argued that even in the 
cases where the STT transformation is well defined it is not particularly useful and that the original variables are required 
anyway in order to interpret the solutions correctly. Furthermore, and as we mentioned above, dealing with the STT approach opens the way to the 
long-standing controversy about which frame (Einstein or Jordan) is the physical one~\cite{frames}. 
This discussion is in some cases semantic and in many others completely ill 
founded and corrupted. In our treatment we don't even need to deal with it at all since we consider the theory directly as it emerges from 
the original action without performing any formal or ``rigorous'' identification or transformation with any other theory, frame or variables.

In the present article we review, in light of our robust approach, the cosmological analysis of some of the apparently least problematic $f(R)$ models, in the sense 
that they seem to pass the cosmological and Solar System tests (although the model of Ref.~\cite{Miranda2009} is still on debate). 
As we will show, our treatment allows to handle the equations 
most as in the case of GR, which in turn, makes possible the use of simple techniques of numerical relativity to monitor the accuracy of the solutions. 

The paper is organized as follows, Section II introduces the $f(R)$ theory, the general equations and the approach under which we will treat them. 
In Section III we present four inequivalent definitions of the ``energy-momentum'' tensor of {\it geometric dark energy} that we have identified 
in the literature and which give rise to three inequivalent equations of state (EOS) used in cosmology. 
Section IV displays in detail the three specific $f(R)$ models that we submit to a cosmological analysis. 
In Sections V and VI we focus on the Friedmann-Robertson-Walker (FRW) cosmology and 
analyze the EOS that arises in the particular $f(R)$ models we treat, as well as the relative abundances of the different 
matter-energy components in order to explicitly show the matter and modified-gravity (geometric dark energy) domination epochs. In order to have 
some insight about the viability of the solutions, we compare the results with the successful GR-$\Lambda CDM$ scenario. We also compute the luminosity 
distance and confront the results with the historic SNIa data~\cite{Riess1998} and the UNION 2 compilation~\cite{Amanullah2010}. 
The age of the Universe that arises from these models is also estimated. Finally, Section VII concludes with a summary and a discussion. 
An Appendix displays the dimensionless form of the cosmological equations used for numerical integration and the numerical test used 
to check the accuracy of our solutions.


\section{$f(R)$ theories, a robust approach}
\label{sec:f(R)}
The MTG that we consider is given by the following action

\begin{equation}
\label{f(R)}
S[g_{ab},{\mbox{\boldmath{$\psi$}}}] =
\!\! \int \!\! \frac{f(R)}{2\kappa} \sqrt{-g} \: d^4 x 
+ S_{\rm matt}[g_{ab}, {\mbox{\boldmath{$\psi$}}}] \; ,
\end{equation}
where  $\kappa \equiv 8\pi G_0$ (we use units where $c=1$), and 
$f(R)$ is an ${\it a priori}$ arbitrary function of the Ricci scalar $R$~\footnote{It is important not to confuse this parametrization 
in the action with the alternative writing 
$f(R)=R + \tilde f(R)$ which is used 
by several authors (c.f. \cite{Hu2007}), and where the tilde is then dropped.}. The first term corresponds to the modified gravity action, while the 
second is the usual action for the matter, where ${\mbox{\boldmath{$\psi$}}}$ represents schematically the matter fields (including both the 
visible and possibly the dark matter).

The field equation arising from  Eq.~(\ref{f(R)}) in the metric approach is
\begin{equation}
\label{fieldeq1}
f_R R_{ab} -\frac{1}{2}fg_{ab} - 
\left(\nabla_a \nabla_b - g_{ab}\Box\right)f_R= \kappa T_{ab}\,\,,
\end{equation}
where $f_R$ indicates $\partial_R f$, $\Box= g^{ab}\nabla_a\nabla_b$ is the covariant D'Alambertian and $T_{ab}$ is the energy-momentum 
tensor of matter which arises from the variation of the matter action in Eq.~(\ref{f(R)}). It is straightforward to 
write the above equation in the following way

\begin{equation}
\label{fieldeq2}
f_R G_{ab} - f_{RR} \nabla_a \nabla_b R - 
 f_{RRR} (\nabla_aR)(\nabla_b R) + g_{ab}\left[\frac{1}{2}\left(Rf_R- f\right)
+ f_{RR} \Box R + f_{RRR} (\nabla R)^2\right]  = \kappa T_{ab}\,\,,
\end{equation}
where $(\nabla R)^2:= g^{ab}(\nabla_aR)(\nabla_b R)$. Taking the trace of this equation yields
\begin{equation}
\label{traceR}
\Box R= \frac{1}{3 f_{RR}}\left[\rule{0mm}{0.4cm}\kappa T - 3 f_{RRR} (\nabla R)^2 + 2f- Rf_R \right]\,\,\,,
\end{equation}
where $T:= T^a_{\,\,a}$. Finally, using Eq.~(\ref{traceR}) in Eq.~(\ref{fieldeq2}) we find
\begin{equation}
\label{fieldeq3}
G_{ab} = \frac{1}{f_R}\Bigl{[} f_{RR} \nabla_a \nabla_b R +
 f_{RRR} (\nabla_aR)(\nabla_b R) - \frac{g_{ab}}{6}\Big{(} Rf_R+ f + 2\kappa T \Big{)} 
+ \kappa T_{ab} \Bigl{]} \; .
\end{equation}
Equations~(\ref{traceR}) and ~(\ref{fieldeq3}) are the basic equations for $f(R)$ theories 
of gravity that we propose, as previously stated in \cite{Jaime2011}, to treat in every application, instead of transforming them to 
STT. Notice that GR with $\Lambda$ is recovered for $f(R)=R-2\Lambda$. It is important 
to stress that even in the treatments where the STT approach is not pursued, in most of the cases 
the term $\Box R$ is not rewritten in the way we do it here, and therefore 
the resulting equations for specific spacetimes turn out to be much more involved. The idea is also to 
rewrite, when possible, all the second order derivatives of $R$ coming from $\nabla_a \nabla_b R$ in Eq.~(\ref{fieldeq3}) 
in terms of lower order derivatives using Eq.~(\ref{traceR}). We shall do that for the cosmological applications 
that we analyze in Sec.~\ref{sec:cosmology}. The systematic approach to perform the 
full first order reduction is through the 3+1 formalism~\cite{Salgado2012}. The method 
proposed here follows the same philosophy of our previous article~\cite{Jaime2011}. 
We also point out that a similar reduction was considered by Seifert~\cite{Seifert2007}, when analyzing the stability of 
several systems under the framework of $f(R)$ theories.

 An important property of these theories is 
the well known fact that not only the total ``energy-momentum'' tensor (i.e. the right-hand-side --r.h.s-- term of Eq.~[\ref{fieldeq3}]) 
is conserved, but also the energy-momentum tensor of matter alone $ T^{ab}$. That is, the field equations imply 
$\nabla_a T^{ab}=0$. This reflects no other but the fact that this kind of theories are metric theories, and 
so, the geodesic equation for test particles holds. The matter equations will take then the same form 
as in GR, and the departure from the latter will occur in the evolution equations that the metric will follow. 

Clearly, we have not modified the fourth-order character of the theory with respect to the metric, since $R$ depends 
on second derivatives of the $g_{ab}$ components and, as we appreciate from the r.h.s of Eq.~(\ref{fieldeq3}), 
there are in addition second derivatives acting on $R$. Nevertheless, the important point here is to promote $R$ as an independent variable, 
and solve the system ~(\ref{traceR}) and ~(\ref{fieldeq3}) as a set of coupled second-order PDE's for $R$ and $g_{ab}$ respectively. 
Of course, the same spirit is used when 
the theory is transformed to an STT counterpart, where instead of $R$ a scalar-field $\chi=f_R$ is defined. However, as we emphasized 
before, we will avoid such a treatment in view of the potential drawbacks that can appear, while in our case everything is as well defined as 
the function $f(R)$ itself. 
In particular, since we shall deal with the Starobinsky~\cite{Starobinsky2007} and Hu--Sawicky~\cite{Hu2007} (hereafter HS1)
models where $f_{RR}$ is not positive definite and for which the scalar-field potential that 
arises in the STT transformation is multivalued, it is then advisable not to pursue that approach.

It is important to stress that following this treatment (i.e. the second-order approach as opposed to the fourth order one for the 
metric alone), the initial data on $R$ and its time derivative are not arbitrary, but subject also to 
a modified Hamiltonian and momentum constraints~\cite{Salgado2012}. 
These important mathematical issues become apparent when formulating the theory as an initial value problem~\cite{Salgado2012}. In the particular case 
of a FRW spacetime, this issue will be reflected in that the initial values for $R$ and $\dot R$ must satisfy 
the equivalent of the Friedman equation for GR, which amounts to the modified Hamiltonian constraint; the momentum constraint being trivially satisfied in this case.
Under this approach, there exists another identity 
which links the Ricci scalar with first and second order derivatives of the metric. This identity 
can provide a consistency test for the numerical integration since, like the initial data constraints, it must be satisfied everywhere in the space-time. 
In particular, in static situations where there is no evolution or the evolution is trivial, this identity can be used to check the self-consistency 
during the numerical integration over the space~\cite{Jaime2011}. A similar situation happens for the FRW case, except that in this case $R$ provides another evolution 
equation for the Hubble expansion. Nevertheless, this equation is consistent with the rest, and so, it provides in practice, a redundant equation, 
since {\it a priori} no extra information can be extracted from it, although it can be used also to fix the initial data (by means of the deceleration 
parameter and the jerk -- see Sec.~\ref{sec:numint}). 
As in the static and spherically symmetric case, this redundancy can be exploited numerically to check the consistency and accuracy of the computer 
codes that we created to solve the equations numerically. 

Finally, we mention what maybe the most important property of $f(R)$ theories as models intending to mimic a cosmological constant at 
present time. First, from Eq.~(\ref{traceR}) a ``potential'' $V(R)= -R f(R)/3 + \int^R f(x) dx$ 
can be defined, so that $V_R(R)=dV(R)/dR =\left(2f-f_{R}R\right)/3$. Then notice that if the matter terms are absent (e.g. outside a compact 
object), almost negligible or small (e.g. at late times of the Universe), then Eq.~(\ref{traceR}) admits as solution $R=R_1=const.$, provided 
$V_R(R_1)=0$ (c.f. Sec.~\ref{sec:models} for specific examples). When this solution is used in (\ref{fieldeq3}) under the approximation $T_{ab}\approx 0 $, this reads 
$G_{ab}= -\Lambda_{\rm eff} g_{ab}$, where $\Lambda_{\rm eff}:= R_1/4$. Of course taking the trace of (\ref{fieldeq3}) is consistent with $R=R_1$. 
Then the solution for the metric must be such that the Ricci scalar is constant. But this is no other than the de Sitter type of solutions. In particular, this 
holds for static and spherically symmetric spacetimes~\cite{Jaime2011,f(R)} and for the FRW cosmology, as we shall see below.
So, in summary, $f(R)$ are able to mimic $\Lambda$ provided that one finds solutions where the matter contribution 
can be neglected, e.g. asymptotically in space or in time, and where the Ricci scalar $R$ approaches a critical point (a maximum or minimum) 
of the ``potential'' $V(R)$. All this relies on $f_{RR}(R_1)\neq 0$, as otherwise the 
conclusions might change. Given that we solve the full equations (\ref{traceR}) and (\ref{fieldeq3}) regardless of any {\it ad hoc} definition of a potential, 
in our case $V(R)$ is merely used as a guiding tool for identifying the critical point associated with the possible asymptotic solution for $R$. 
Then, only for such guiding purposes and for models where $f_{RR}$ is positive definite in general or that $0<f_{RR}<\infty$ in the regions where the solutions 
take place, it is irrelevant to include the term $f_{RR}$ in $V(R)$, since we will find the same critical points. 
Nevertheless, when those conditions are not fulfilled (i.e. when 
$f_{RR}(R_1)=0$ or when $f_{RR}(R_1)\rightarrow \infty$), it is then advisable to include the term $f_{RR}$ in the potential, otherwise one could ``miss'' a critical 
point (see Sec.~\ref{sec:models} for a further discussion). For the $f(R)$ models that we analyze below, some of which have a non positive definite $f_{RR}$ 
(see Sec.~\ref{sec:models}) it turns however, that the cosmological solutions presented in Sec.~\ref{sec:numerics}, 
never reach the point(s) where $f_{RR}\leq 0$ nor where $f_{RR}$ diverges. Therefore, there exists viable cosmological models that are free of 
any ``pathologies'' of this sort.

\section{The ``Energy-Momentum Tensor'' of $f(R)$}
\label{sec:EMT}
Very often it turns to be convenient and helpful to write the field equations of alternative (metric) theories of gravity as the Einstein field equations with an 
effective (total) energy-momentum tensor (EMT) that contains all the modifications which are associated with the new theory and the EMT of matter itself. 
In cosmology this rearrangement of the equations can be useful as one tries to identify the contributions of the modifications of gravity within the total 
EMT as though they represented some kind of {\it geometric dark energy}, and so it will be dubbed. This can be specially advantageous since one can define then 
an EOS associated with such dark energy and compare it with the $\Lambda CDM$ model. The only problem with this construction is that, 
given such total EMT, there is no canonical way to perform the separation between the matter and the geometric dark energy contribution, 
and thus, very often different authors introduce different definitions for the EMT of geometric dark energy. The reason is that 
even the matter energy-momentum tensor in Eq.~(\ref{fieldeq3}) appears to be multiplied by a factor $f_R^{-1}$, 
and so, it would seem {\it a priori} difficult to unambiguously define which part of the field equations belongs purely to the matter terms and which 
corresponds to the geometry (c.f. the remarks at the end of Sec.~IIA of Ref.~\cite{Sotiriou2010}). 

In the following we provide four inequivalent definitions of the EMT of geometric dark energy, that when applied to cosmology, lead to three inequivalent 
ways of defining its corresponding EOS. In our opinion, the fact that these various definitions have been considered in the literature 
without even identifying them as {\it inequivalent}, has added a great amount of confusion to the subject. We hope that 
by clearly exposing these definitions we help to clarify this matter.

{\bf Recipe I:} First, define $\kappa\,T_{ab}^{\rm tot}$ as 
the r.h.s. of Eq.~(\ref{fieldeq3}). Then, define the EMT of the geometric dark energy as $T_{ab}^{X}:= T_{ab}^{\rm tot} - T_{ab}$~\footnote{In several articles $T_{ab}^{X}$ 
is denoted by $T_{ab}^{\rm eff}$ instead. Nevertheless, sometimes it 
goes beyond purely notation and different meanings are to be understood in both symbols (c.f. Recipe III).}. 
By construction, $T_{ab}^{X}$ is conserved, for $T_{ab}^{\rm tot}$ is conserved by the Bianchi identities, and as mentioned in Sec.~II, the energy-momentum 
tensor $T_{ab}$ of matter alone turns to be also conserved 
(c.f. Appendix A of Ref.~\cite{Hu2007b} for similar considerations and also Ref.~\cite{Starobinsky2007} for further reflexions). 
Moreover, in the GR case $f(R)=R$, $T_{ab}^{\rm tot}\equiv T_{ab}$, which in turn leads to $T_{ab}^{X}\equiv 0$, even in the 
presence of matter, as one can verify from Eq.~(\ref{EMTX1a}) below. We conclude that $T_{ab}^{X}$ includes a non trivial contribution only when GR 
is modified, and thus captures the idea about {\it the} energy-momentum content of the geometric dark energy. 
The explicit form of $T_{ab}^{X}$ is as follows:
\begin{equation}
\label{EMTX1a}
 T_{ab}^{X} := \frac{1}{\kappa f_R}\Bigl{[} f_{RR} \nabla_a \nabla_b R + f_{RRR} (\nabla_aR)(\nabla_b R) 
- \frac{g_{ab}}{6}\Big{(} Rf_R+ f + 2\kappa T \Big{)} + \kappa T_{ab}\left(1- f_R\right) \Bigl{]} \,.
\end{equation}
As we mentioned above, clearly for $f(R)=R$, $T_{ab}^{X}\equiv 0$, and for $f(R)= R - 2\Lambda$, 
$T_{ab}^{X}= - \Lambda g_{ab}/\kappa$, where we have used $R= -\kappa\,T^{\rm tot}$. 

It may perhaps seem awkward to see the EMT of matter $T_{ab}$ appearing in the definition of the EMT of geometric dark energy Eq.~(\ref{EMTX1a}). Nonetheless there is 
a simple way to rewrite $T_{ab}^{X}$ in terms of purely geometric quantities. Adding $G_{ab}$ to both sides of Eq.~(\ref{fieldeq2}) and 
defining $T_{ab}^{X}$ such that $ G_{ab}=  \kappa\,T_{ab}^{X} + \kappa\,T_{ab}$, we obtain
\begin{equation}
\label{EMTX1b}
\kappa\,T_{ab}^{X} = G_{ab}(1-f_R) + f_{RR} \nabla_a \nabla_b R + f_{RRR} (\nabla_aR)(\nabla_b R) - g_{ab}\left[\frac{1}{2}\left(Rf_R- f\right)
+ f_{RR} \Box R + f_{RRR} (\nabla R)^2\right]  \,.
\end{equation}
Alternatively we can replace $\Box R$ using Eq.~(\ref{traceR}), and get
\begin{equation}
\label{EMTX1c}
\kappa\,T_{ab}^{X} = G_{ab}(1-f_R) + f_{RR} \nabla_a \nabla_b R + f_{RRR} (\nabla_aR)(\nabla_b R) 
- \frac{g_{ab}}{6}\Big{(} Rf_R+ f + 2\kappa T \Big{)}  \,,
\end{equation}
which of course can be recovered by using $\kappa T_{ab}= G_{ab}- \kappa T_{ab}^{X}$ in Eq.~(\ref{EMTX1a}) and then solving for 
$T_{ab}^{X}$. Like in Eq.~(\ref{EMTX1a}) the matter contribution appears also in Eq.~(\ref{EMTX1c}) via the trace $T=T^a_{\,\,a}$ of the EMT of matter.

We have then arrived to three equivalent ways of writing $T_{ab}^{X}$ which are given by Eqs.~(\ref{EMTX1a})$-$(\ref{EMTX1c}), 
except that now in Eq.~(\ref{EMTX1b}) the matter contribution does not appear explicitly. 
Expression ~(\ref{EMTX1b}) corresponds exactly to the EMT of geometric dark energy considered in 
Refs.~\cite{Starobinsky2007,Motohashi2010,Motohashi2011a,Motohashi2011b}
\footnote{In Ref.~\cite{Starobinsky2007} the signature $(+,-,-,-)$ is used along with a different sign convention for its $T_{ab}^{DE}$.}.
Following the line of thought that we have proposed in Sec.~II, and 
for the cosmological applications, it will be better for us to work with Eq.~(\ref{EMTX1a}) rather than with the other two equivalent expressions.
\bigskip

{\bf Recipe II:} In order to obtain the second proposal for the EMT we follow a prescription similar to Recipe I, but we write the EMT of the 
geometric dark energy as the following linear combination 
$T_{ab}^{II\,,\,X}(A):=  A\,T_{ab}^{\rm tot} -T_{ab}=  A\kappa^{-1} G_{ab} - T_{ab} $ where $A$ is a constant. This is a rather {\it ad-hoc} generalization of $T_{ab}^{X}$ which 
in our opinion has no deep motivation. By the same arguments given above, this EMT will also be conserved. Explicitly it reads
\begin{equation}
\label{EMTX2a}
\kappa\,T_{ab}^{II\,,\,X}(A) := G_{ab}(A-f_R) + f_{RR} \nabla_a \nabla_b R + f_{RRR} (\nabla_aR)(\nabla_b R) 
- g_{ab}\left[\frac{1}{2}\left(Rf_R- f\right) + f_{RR} \Box R + f_{RRR} (\nabla R)^2\right]  \,,
\end{equation}
so that Eq.~(\ref{fieldeq2}) reads $AG_{ab}= \kappa\,T_{ab}^{II\,,\,X} + \kappa\,T_{ab}$. Thus, $T_{ab}^{X}= T_{ab}^{II\,,\,X}(1)$ , that is, 
$T_{ab}^{II\,,\,X}$ and $T_{ab}^{X}$ coincide if and only if $A=1$. Several authors~\cite{Amendola2007b,Tsujikawa2007,Amendola2008,Gannouji2009} considered this recipe in the 
cosmological context and set $A= f_R^0$ ($F_0$ in their notation; the knot indicating today's value). Therefore when $f_R^0$ is not taken as unit, as may be often the case, 
the EOS obtained from Recipes I and II are not equivalent (see Secs.~\ref{sec:EOS}, and \ref{sec:numerics} ). Actually, the EOS for this recipe has the  
unappealing feature that can be divergent in several cases~\cite{Amendola2008} as we will show in 
Sec.~\ref{sec:numerics} (c.f. Fig.~\ref{fig:wx2-St-Mir}). This divergence is due to the fact that in a FRW cosmology the energy-density associated with this EMT becomes 
zero at some redshift (c.f. Figs.~\ref{fig:rhoXMir} and \ref{fig:rhoXSt}), a feature that had already been remarked by Starobinsky~\cite{Starobinsky2007}.

Using Eq.~(\ref{fieldeq3}) one can also write
\begin{equation}
\label{EMTX2b}
T_{ab}^{II\,,\,X}(A) = \frac{A}{\kappa f_R}\Bigl{[} f_{RR} \nabla_a \nabla_b R + f_{RRR} (\nabla_aR)(\nabla_b R) 
- \frac{g_{ab}}{6}\Big{(} Rf_R+ f + 2\kappa T \Big{)} + \kappa T_{ab}\left(1- \frac{f_R}{A}\right) \Bigl{]} \,.
\end{equation}
\bigskip

{\bf Recipe III:} The EMT defined in this recipe arises from Eq.~(\ref{fieldeq2}) or Eq.~(\ref{fieldeq3}) by simply identifying
\begin{eqnarray}
\label{EMTX3}
\kappa\,T_{ab}^{III\,,\,X} &:=& f_{RR} \nabla_a \nabla_b R + f_{RRR} (\nabla_aR)(\nabla_b R) 
-  g_{ab}\left[\frac{1}{2}\left(Rf_R- f\right) + f_{RR} \Box R + f_{RRR} (\nabla R)^2\right] \nonumber \\
&=&  f_{RR} \nabla_a \nabla_b R + f_{RRR} (\nabla_aR)(\nabla_b R) 
- \frac{g_{ab}}{6}\Big{(} Rf_R+ f + 2\kappa T \Big{)} \; .
\end{eqnarray}
so that Eqs.~(\ref{fieldeq2}) and (\ref{fieldeq3}) read $G_{ab}= \frac{\kappa}{f_R}\left(\,T_{ab}^{III\,,\,X} + T_{ab}\right)$. This EMT was considered 
by Sotiriou \& Faraoni~\cite{Sotiriou2010} (denoted $\,T_{\mu\nu}^{(eff)}$ by them). It has the unpleasant feature that is not conserved, not 
even in the absence of matter, by the fact that $f_R^{-1}$ is not included as a factor in the r.h.s of Eq.~(\ref{EMTX3}). 
\bigskip

{\bf Recipe IV:} 
In this case the EMT is defined by $T_{ab}^{IV\,,\,X} := f_R^{-1} T_{ab}^{III\,,\,X}$ so that Eqs.~(\ref{fieldeq2}) and (\ref{fieldeq3}) read 
$G_{ab}= \kappa T_{ab}^{IV\,,\,X} + \kappa f_R^{-1} T_{ab}$. Like $T_{ab}^{III\,,\,X}$, this EMT does not conserve either since in the presence of matter 
$\nabla^a T_{ab}^{IV\,,\,X} = -T_{ab}\nabla^a (f_R^{-1}) \neq 0$, however, unlike $T_{ab}^{III\,,\,X}$, it is conserved in the absence of matter, 
as can be seen from the previous equation. As we shall discuss in Sec.~\ref{sec:EOS}, several authors have obtained an EOS from the fourth 
recipe when applied to cosmology. In such scenario the previous equation will imply that the conservation equation associated with the geometric dark energy 
will contain a source term given by $-T_{tt}\nabla^t (f_R^{-1})= \rho \,d (f_R^{-1})/dt$ (where $t$ is the cosmic time) and thus such term will depend on the 
total matter density (i.e. radiation plus baryon plus dark matter densities). 
Incidentally, even if $\,T_{ab}^{IV\,,\,X}\neq T_{ab}^{III\,,\,X}$ the EOS associated with both EMT will coincide as 
the factor $f_R^{-1}$ cancels when taking the ratio of pressure over energy-density for the $X$--component.

When using Eq.~(\ref{fieldeq3}), $\,T_{ab}^{IV\,,\,X}$ can be written as follows 
\begin{equation}
\label{EMTX4}
T_{ab}^{IV\,,\,X} = \frac{1}{\kappa f_R}\Bigl{[} f_{RR} \nabla_a \nabla_b R + f_{RRR} (\nabla_aR)(\nabla_b R)  
- \frac{g_{ab}}{6}\Big{(} Rf_R+ f + 2\kappa T \Big{)} \Bigl{]} \,.
\end{equation}

In our opinion Recipe I is the simplest, the best motivated EMT and the one that has the nicest features. 
In Sec.~\ref{sec:EOS} we shall derive the different EOS that arise from Recipes I--IV.


\section{$f(R)$ models}
\label{sec:models}
We have selected three $f(R)$ models which have been analyzed carefully in recent years. 
First we shall present the models by reviewing some general features, and then test their cosmological viability using the approach that we will discuss 
in the next sections where we will compare our findings with previous results. In particular, we confront them in the light of the $\Lambda CDM$ paradigm.

\subsection{MJW model}
\begin{figure}
\includegraphics[width=9cm]{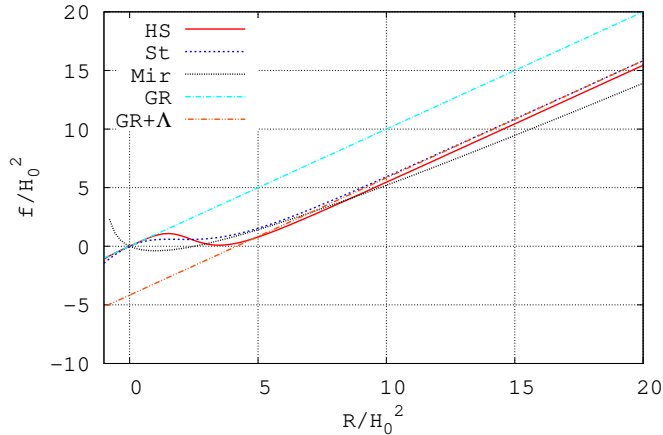}
\caption{$f(R)$ models considered here (see the main text for the analytical expressions). For reference the general relativity (GR) case 
$f(R)_{\rm GR}=R$ is also depicted as well as a GR$\Lambda$ model $f(R)_{\rm GR\Lambda}=R-2\Lambda$ with $\Lambda= 2.08 H_0^2$. 
Notice that for sufficiently large $R/H_0^2$, the Starobinsky and the Hu--Sawicky models can be approximated by their respective 
$f(R)_{\rm approx}:= R - 2\Lambda_{\rm eff}^\infty$. Such models behave as $f(R)_{\rm GR\Lambda}$ 
(i.e. almost straight lines) with an effective positive cosmological constant 
$\Lambda_{\rm eff}^\infty$ given by the value $-f(0)_{\rm approx}/2$.}
\label{fig:F(R)'s}
\end{figure}

\begin{figure}[ht]
\includegraphics[width=9cm]{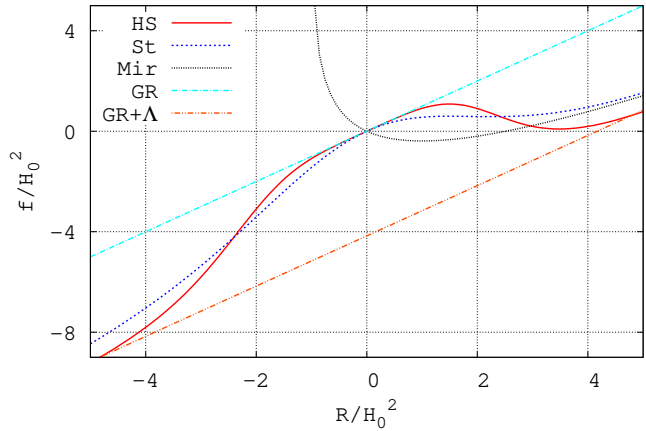}
\caption{Same as Fig.~\ref{fig:F(R)'s} for smaller $R/H_0^2$.}
\label{fig:F(R)zoom}
\end{figure}

\begin{figure}[ht]
\includegraphics[width=9cm]{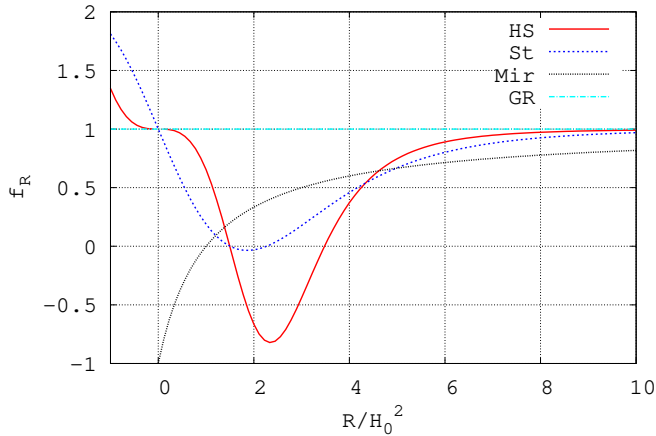}
\caption{The derivative $f_R:= df/dR$ for the $f(R)$ models depicted in Fig.~\ref{fig:F(R)'s}. This quantity turns to be positive during the 
cosmic evolution.}
\label{fig:fR}
\end{figure}

\begin{figure}[ht]
\includegraphics[width=9cm]{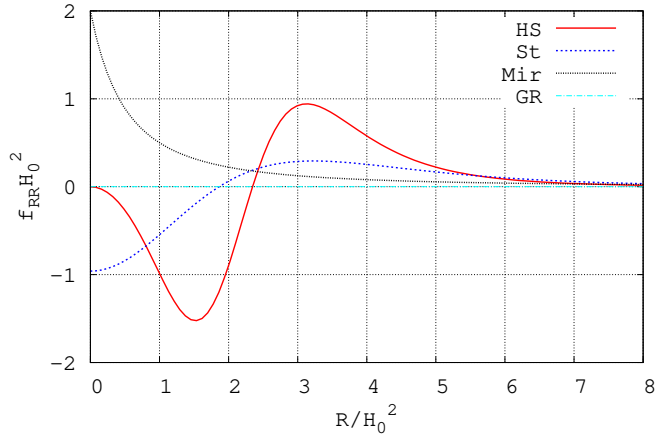}
\caption{The second derivative $f_{RR}:= d^2f/dR^2$ for the $f(R)$ models depicted in Fig.~\ref{fig:F(R)'s}. This quantity is 
positive definite in the MJW model and turns to be positive during the cosmic evolution for the Starobinsky and Hu--Sawicky models considered 
here.}
\label{fig:fRR}
\end{figure}

This model was proposed by Miranda {\it et al.}~\cite{Miranda2009}:
\begin{equation}
\label{f(R)MJW}
f(R)_{\rm MJW}=R-\beta R_{*}{\rm ln}\left( 1+\frac{R}{R_{*}}\right)\,\,\,,
\end{equation}
where $\beta$ and $R_{*}$ are free positive parameters, and it is well defined provided $R/R_{*}> -1$. In this work we use $\beta=2$ and $R_{*}=  \sigma_* H_0^2$, 
where $H_0^2$ is the Hubble constant today, and $\sigma_*$ is a dimensionless parameter that can be adjusted in order to best fit the model to the 
cosmological observations. In the numerical analysis presented in Sec.~\ref{sec:numerics} we take $\sigma_*=1$, which is not necessarily the best 
fit. Figures~\ref{fig:F(R)'s} and~\ref{fig:F(R)zoom} depict 
the behavior of this model. The $f-$curvature $f_{RR}= \beta R_*^{-1}/(R/R_*+1)^2$ (where we have dropped the label MJW) 
is positive definite, provided $\beta>0$ and $R_*>0$, and becomes small for large $R$ (see Fig.~\ref{fig:fRR}). 
In the domain where $f(R)$ is defined, the derivative $f_R$ can be negative or zero in the range $ -1 < R/R_* \leq \beta -1$, 
i.e., $ -1 < R/R_* \leq 1$ for the specific value $\beta=2$. However, as we shall see in Sec.~\ref{sec:numerics}, 
it turns that $f_R>0$ for the cosmological model that we analyzed, as the inequality $R/H_0^2>1$ largely holds during the cosmic evolution 
(c.f. Figs.~\ref{fig:fR} and \ref{fig:RMir} ), avoiding in this way any possible drawback associated with the values $f_R\leq 0$. 
Moreover, since $R/H_0^2>1$, the pole at $R/R_*=-1$, where $f_{RR}\rightarrow \infty$ and where $f(R)$ is no longer defined, 
is never reached (c.f. Fig.~\ref{fig:fRR}), therefore, in this case it is unimportant to include $f_{RR}$ in the potential $V(R)$ that we introduce below.

\begin{figure}[ht]
\includegraphics[width=9cm]{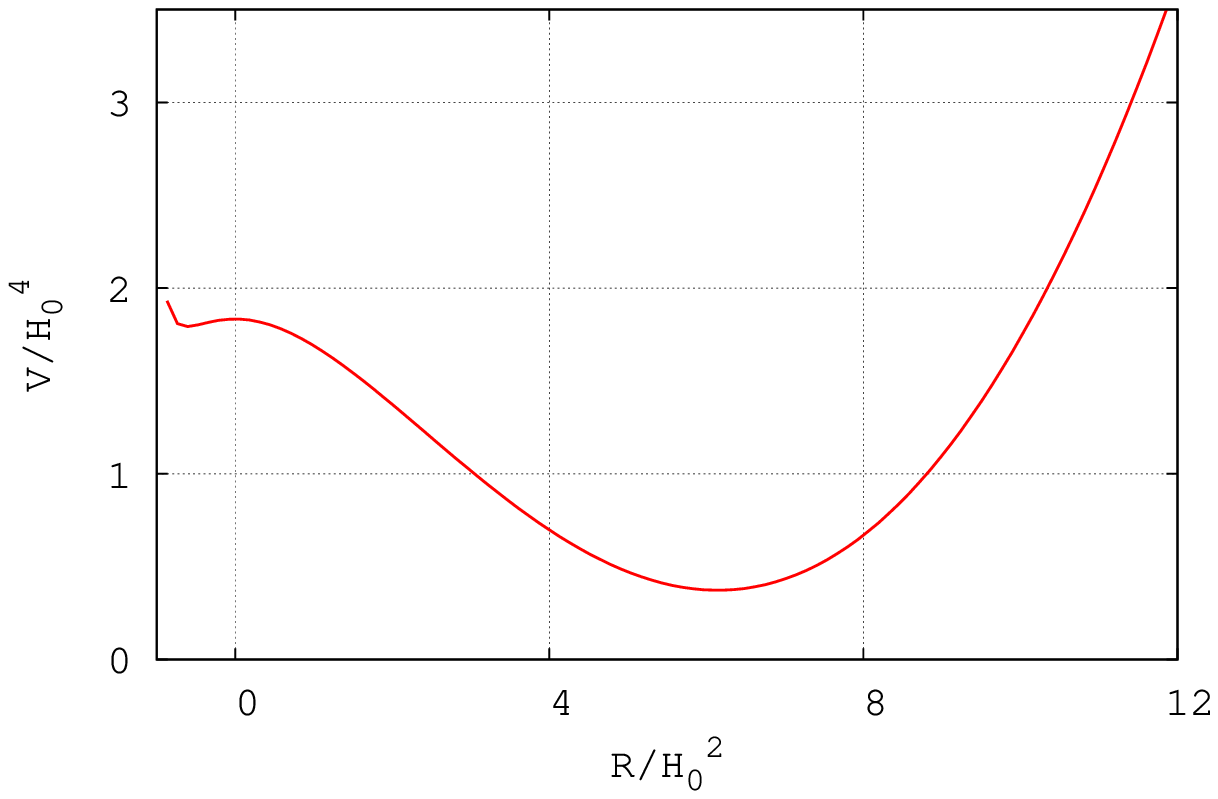}
\caption{The potential $V(R)$ associated with the MJW model Eq.(\ref{f(R)MJW}) with $\beta = 2$, $R_*=H_0^2$.}
\label{VMiranda}
\end{figure}

As we emphasized in the Introduction, the transformation from $f(R)$ gravity to the corresponding STT can generate several pathologies because 
the potential $U(\chi)$ which is associated with the scalar field $\chi= f_R$ may be ill defined (notably when $f_{RR}$ is not positive 
definite since then $\chi(R)$ cannot be inverted in the original domain of $R$). Notwithstanding, the model proposed 
by Miranda {\it et al.}~\cite{Miranda2009} is free from those pathologies since in this case $f_{RR}>0$, unlike the Starobinsky and Hu-Sawicky models discussed below. 
The potential $V(R)$ introduced in Sec.~\ref{sec:f(R)}, and which is different from $U(\chi)$~\footnote{The relationship between both potentials is 
given by $d \bar U (R(\chi))/d\chi= \left(2f-f_{R}R\right)/3 = dV(R)/dR$, where $U(\chi):= \bar U(R(\chi))$. 
Thus, $d \bar U(R) /dR = f_{RR} dV(R)/dR$, where we used $d\chi/dR=f_{RR}$. If we include the term $f_{RR}$ in the definition of an alternative potential
$\tilde V(R)= \left(2f-f_{R}R\right)/(3 f_{RR})$, then the relationship between $\bar U(R(\chi))$ and $\tilde V(R)$
 $d\bar U(R(\chi))/dR = (f_{RR})^2 d\tilde V(R)/dR$.}, 
is given as follows for this model $V(R) =\frac{R_*^2}{6}\Big{\{} (1+\tilde R)(\tilde R + 6\beta-1) 
 -2\beta(3+2\tilde R){\rm ln}(1+ \tilde R) \Big{\}}$, where $\tilde R= R/R_*$. Figure~\ref{VMiranda} depicts this potential where the critical points 
(maxima or minima) at $R_1$ correspond to a possible trivial solution $R= R_1$ in vacuum (i.e. de Sitter point associated with $R_1>0$), 
and are also the {\rm points}, notably the minimum, that should be reached asymptotically, 
(in time for cosmology and in space for compact objects) when a non trivial solution for $R$ 
approaches its asymptotic value in spacetime regions where the matter becomes practically absent or very diluted. The global minimum at $R_1\approx 6.14 H_0^2$ 
is the actual de Sitter point reached in the cosmic evolution (c.f. Fig.~\ref{fig:RMir}). Similar considerations regarding the critical points 
will also apply for the two models discussed below.

Miranda {\it et al.}~\cite{Miranda2009} showed that such a model is consistent with a FRW cosmology and that singularities do not appear 
in idealized compact objects. Regarding these latter, we confirmed their findings using our approach~\cite{Jaime2011}. 
As concerns the cosmological part, they specifically showed that their model is able to reproduce the matter-dominated epoch as well as the accelerated phase 
that are required to explain several observations. This is something that we confirm here (see Sec.~\ref{sec:numerics}). 
This model was previously criticized on the grounds that, at the perturbation level, it seems to be unable to reproduce the observed matter power spectrum and 
that it may also be inconsistent with the local gravity constraints imposed by 
the Solar System experiments~\cite{delaCruz2009}. Miranda {\it et al.}~\cite {Miranda2009b} replied that the arguments put forward in Ref.~\cite{delaCruz2009}
to discard the model are not strong enough and that a careful reexamination is needed. Furthermore, Thongkool {\it et al.}~\cite{Thongkool2009} 
have also argued that this model is unable to satisfy the thin shell condition that is needed to produce a successful chameleon mechanism, being this presumably the 
only way to satisfy the Solar System constraints in $f(R)$ theories.

\subsection{Starobinsky model}
\begin{figure}[ht]
\includegraphics[width=9cm]{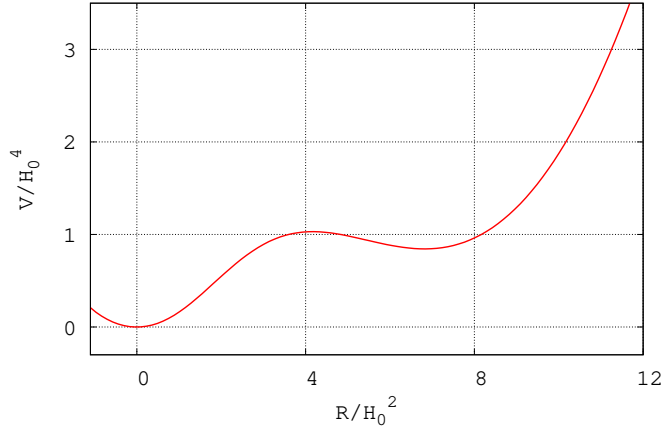}
\caption{Same as Fig.~\ref{VMiranda} for the Starobinsky model Eq.~(\ref{f(R)St}) with $q=2$,  $\lambda =1$ and $R_s\approx 4.17 H_0^2$ .}
\label{VStarobinsky}
\end{figure}

The model is defined by the following function~\cite{Starobinsky2007},
\begin{equation}
\label{f(R)St}
f(R)_{\rm S}=R+\lambda R_{S}\left[ \left( 1+\frac{R^2}{R^2_{S}}\right)^{-q}-1\right]\,\,,
\end{equation} 
with $q, \lambda$ positive parameters and $R_S$ is another parameter playing the same roll as $R_{*}$ in the previous model; we assume 
$R_S= \sigma_S H_0^2$, and take $\sigma_S\approx 4.17$, $q=2$ and $\lambda=1$ as in Ref.~\cite{Motohashi2011a}. 
Both models, this and the previous one, 
have the property that, in vacuum, admit solutions with $R\equiv 0$, like GR in vacuum (which includes the asymptotically flat solutions), 
unlike some models which contain a term $\sim R^{-1}$ in the Lagrangian. 
It is by virtue of this property that Starobinsky entitled his model in terms of a ``disappearing cosmological constant in $f(R)$ gravity''. On the other hand, 
in the high curvature regime where $|R|\gg R_S$, the model yields $f(R)\approx R- \lambda R_S$, and thus it acquires an effective 
cosmological constant $\Lambda_{\rm eff}^{\infty}:= \lambda R_S/2$ (see Figs.~\ref{fig:F(R)'s} and~\ref{fig:F(R)zoom}), 
which is non-negligible provided $\lambda\gg 1$. 
In this model, $f_{RR}=0$ at $R= \pm R_S/\sqrt{2q +1}$, thus, $f_{RR}$ is not positive definite (see Fig.~\ref{fig:fRR}). The fact that 
this quantity appears in the denominator in Eq.~(\ref{traceR}) indicates that a careful examination of the equations is required at those particular points, notably 
the positive one. Starobinsky himself~\cite{Starobinsky2007}, who dubbed such points {\it weak singularities}, has stressed the need of a close analysis of 
the solutions there. Nevertheless, we have not reached any of those weak singularities during the cosmological 
evolution as $R\ge R_S/\sqrt{5}\approx 1.86 H_0^2$  (c.f. Figs.~\ref{fig:fRR} and \ref{fig:RSt}). 
Figure ~\ref{VStarobinsky} shows the potential
\begin{equation}
V(R)=\frac{1}{6}\left(R^{2}-\lambda R R_S\frac{4R^{4}+5R^{2}R_{S}^{2}+3R_{S}^{4}}{(R^{2}+R_{S}^2)^{2}}\right) +\frac{\lambda R_S^{2}}{2} {\rm arctan}(R/R_S)\,\,\,.
\end{equation}
For $R\geq 0$, there is a global minimum at $R=0$, a maximum at $R\approx 4.17 H_0^2$ and a local minimum 
at $R_1\approx 6.82 H_0^2$ which corresponds to the actual de Sitter point where the cosmological solution settles in future cosmic time 
(c.f. Fig.~\ref{fig:RSt}). Like in the previous model, the cosmological solution in the range we explored is such that $f_{RR}$ and $f_R$ 
are always positive. It is somehow remarkable that $\Lambda_{\rm eff}^{\infty}\approx 2.09 H_0^2$ is close to the actual effective cosmological 
constant $\Lambda_{\rm eff}= R_1/4\approx 1.71 H_0^2$, even though the minimum is reached only for $R_1\approx 1.67 R_S$ which one would think 
is not yet in the regime  $R\gg R_S$.

For the model with $q=1$ and $\lambda=1.56$, we found previously static and spherically symmetric solutions that 
represent idealized compact objects with an asymptotically de Sitter behavior that 
was reached at the local minimum of $V(R)$~\cite{Jaime2011}. In this regard, it is important to mention that we did 
not hit any of those {\it weak singularities}, notably the positive value) as the solution interpolates monotonically between $R\approx 2.37 R_S$ at the center of the 
object and $R\approx 1.98 R_S$ asymptotically, which, as mentioned, corresponds to a de Sitter point~\cite{Jaime2011}. However, it is fair to say that in that work 
we did not explore sufficiently the space of solutions and different values of the parameters. Moreover, we only used incompressible fluids. 
We plan to extend this analysis elsewhere. 

Starobinsky's model can satisfy the conditions imposed by several cosmological observations~\cite{Amendola2008,Tsujikawa2008}. 
For instance, it is able to produce an adequate matter epoch prior to the accelerated era, 
unlike several unsuccessful models (see Refs.~\cite{Amendola2007a,Amendola2007b,Amendola2007c} for a thorough analysis). Moreover, the Solar System tests can be 
successfully passed by this model (e.g. taking $n \geq 2$)~\cite{Starobinsky2007}. However it leads to ill defined 
potentials when transformed to a STT because $f_{RR}$ is not positive definite, and therefore the analyzes that rely on such an approach rise serious doubts about 
their soundness~\cite{Frolov2008,Dev2008,Capozziello2009,Kobayashi2008,Kobayashi2009,Babichev2009,Upadhye2009}, even if some of their conclusions turn to be true 
(c.f. Ref.~\cite{Jaime2011}).

\subsection{Hu--Sawiky model}
The $f(R)$ for this model is given as follows~\cite{Hu2007}:
\begin{equation}
\label{f(R)HS}
f(R)=R-m^2\frac{c_{1}(R/m^2)^n}{c_{2}(R/m^2)^n+1}\,\,,
\end{equation} 
with $n>0$ and $m^2$, $c_1$ and $c_2$ are parameters of the model. Like the previous two models, $m^2$ plays the same roll as $R_*$ and $R_S$. 
According to HS1, the constant $m^2$ is fixed from the length scales of the Universe 
(it has the same units as $R$, so $1/m$ has units of length),
\begin{equation}
m^2\sim \frac{\kappa^2 {\rho_0}}{3} \,\,,
\end{equation}
where $\rho_0$ is the average density of the Universe today. The numerical value that we assume and which is similar to the value taken in HS1 is 
$m^2\approx 0.24 H_0^2$. The constants $c_1$ and $c_2$ are dimensionless parameters that can be fixed by demanding that this model mimics as close as 
possible the $\Lambda CDM$ scenario. In particular their values are chosen for the model to match the current values 
$\,^{\Lambda CDM}\,\!\Omega_{\rm bar+DM}^0  \approx 0.24$ and 
$\Omega_\Lambda^0 \approx 0.76$ (see Sections~\ref{sec:cosmology} and \ref{sec:numerics} for definitions), 
where the upper-left index $\Lambda CDM$ is used to distinguish from the corresponding matter content predicted by $f(R)$ theories, which in general will not be 
exactly the same. The relative density $\Omega_\Lambda^0$ will be replaced by a suitably defined energy-density of the 
geometric dark energy. Following HS1 one fixes such constants from
\begin{equation}
\frac{c_1}{c_2}= \frac{6\Omega_\Lambda^0}{\,^{\Lambda CDM}\,\!\Omega_{\rm bar+DM}^0}\,\,,
\end{equation} 
and
\begin{equation}
f_{R}^{0}-1 =-n\frac{c_{1}}{c_{2}^{2}}\left(\frac{12}{\,\,^{\Lambda CDM}\,\!\Omega_{\rm bar+DM}^0}-9\right)^{-n-1}\,\,.
\end{equation}
In this paper we take $n=4$ and~\footnote{The reader is urged to remember our definition of $f(R)$ which contains the Ricci scalar unlike the 
HS1 convention. Therefore $f_R^0= 1 +\,^{\rm HS}f_R^0$. So our value $f_R^0=0.99$ corresponds to $^{\rm HS}f_R^0= -0.01$.} $f_R^0=0.99$ 
so that the specific values for $c_1$ and $c_2$ are $c_1\approx 1.25 \times 10^{-3}$, 
$c_2\approx 6.56 \times 10^{-5}$. Figures~\ref{fig:F(R)'s} and \ref{fig:F(R)zoom} depict this model and show that 
in the regime $R\gg m^2$, $f(R)\approx R- c_1 m^2/c_2$, where now $\Lambda_{\rm eff}^\infty= c_1 m^2/(2c_2)$ is not negligible 
provided $c_1\gg c_2$. Like in the Starobinsky case, the Hu--Sawicky model has the property that $f_R$ and $f_{RR}$ are positive 
during the cosmic evolution (c.f. Figs.~\ref{fig:fR},~\ref{fig:fRR} and~\ref{fig:RHu}), even though they are not positive definite.

\begin{figure}[ht]
\includegraphics[width=9cm]{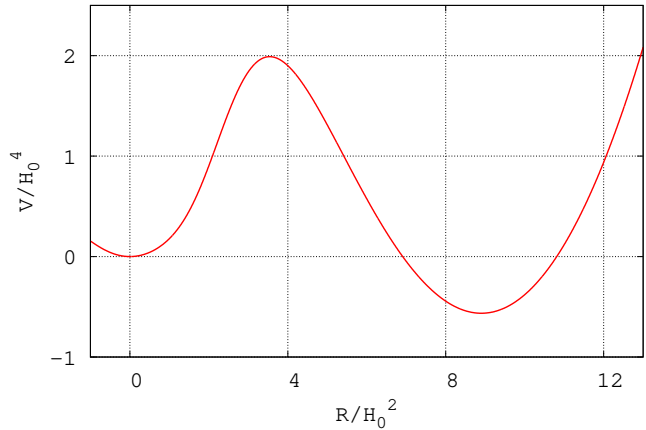}
\caption{Same as Fig.~\ref{VMiranda} for the Hu--Sawicky model Eq.~(\ref{f(R)HS}). Here $n=4$, $m^2= 0.24 H-0^2$, $c_1\approx 1.25 \times 10^{-3}$, and 
$c_2\approx 6.56 \times 10^{-5}$.}
\label{VHu}
\end{figure}

The potential $V(R)$ is shown in Figure~\ref{VHu}. The analytical expression for $V(R)$ in this 
case is not very enlightening (it is given in terms of a hypergeometric function) and thus we do not write it explicitly here, but 
as one can appreciate, its structure is similar to the potential of 
the Starobinsky model, except that the global minimum at $R_1\approx 8.9 H_0^2$ corresponds to the actual de Sitter point found in the cosmic evolution 
(c.f. Fig.~\ref{fig:RHu}). It is worth stressing that $\Lambda_{\rm eff}^\infty \approx 2.29 H_0^2$ while $\Lambda_{\rm eff}= R_1/4\approx 2.23 H_0^2$. 
Those values are very close to each other since in this case $R_1\approx 37.08 m^2$, which is already in the regime $R\gg m^2$.  

As shown explicitly in HS1, the authors constructed a spherically symmetric static solution that 
represents the spacetime outside the Sun which passes the Solar System tests via a chameleon mechanism, as it provides a post-Newtonian parameter 
$\gamma\approx 1$. Furthermore, like the Starobinsky model, it is also consistent with the required matter dominated epoch prior to the accelerated expansion era 
as we shall see in Section~\ref{sec:numerics}. Notice that the Hu-Sawicky model with $n=2$ is essentially the same as the 
Starobinsky model with $q=1$ modulo a redefinition of their parameters.

The Hu--Sawicky and Starobinsky $f(R)$ models are perhaps the most tested models so far, which includes confrontation with CBR, SNIa, baryon acoustic oscillations 
and gravitational lensing among others~\cite{HS&Staromodel,Martinelli2009,Martinelli2012}~\footnote{In several of theses references, generic $f(R)$ 
models are tested using a parametrization that characterizes the deviations relative to GR.}.

\section{Cosmology in $f(R)$}
\label{sec:cosmology}

We focus now on homogeneous and isotropic space-times which are relevant for cosmology and which are described by the FRW metric,
\begin{equation}
\label{SSmetric}
ds^2 = - dt^2  + a^2(t) \left[ \frac{dr^2}{1-k r^2} + r^2 \left(d\theta^2 + \sin^2\theta d\varphi^2\right)\right]\,\,\,, 
\end{equation}
where $k=\pm 1,0$. 

From Eq.~(\ref{traceR}) we find
 \begin{equation}
\label{traceRt}
  \ddot R = -3H \dot R - \frac{1}{3 f_{RR}}\left[3f_{RRR} \dot R^2 + 2f- f_R R + \kappa T \right] \,\,\, ,
\end{equation}
where $\dot\,\,= d/dt$. Like in GR, the diagonal spacetime components of Eq.~(\ref{fieldeq3}) will provide the remaining field equations. However, note that 
the $t-t$ component of Eq.~(\ref{fieldeq3}) will contain a term like $\ddot R$. This term will be rewritten in terms of lower order derivatives 
using Eq.~(\ref{traceRt}). The final expression for the equations which have the desired form of a 
Cauchy initial-value problem, subject to the initial data constraint is:

\begin{equation}
\label{Friedmod}
H^2 + \frac{k}{a^2} + \frac{1}{f_R}\left[ f_{RR} H \dot R -\frac{1}{6}\left( f_R R- f\right)\right]= \frac{-\kappa T^t_{\,\,t}}{3f_R}\,\,,
\end{equation}
\begin{equation}
\label{accmod}
\dot H = -H^2 + \frac{1}{f_R}\left(f_{RR} H \dot R + \frac{f}{6} + \frac{\kappa T^t_{\,\,t}}{3} \right) \,\,\,\,,
\end{equation}
where 
\begin{equation}
\label{Hubble}
H= \dot a/a \,\,\,,
\end{equation}
 is the Hubble expansion.

Equation~(\ref{Friedmod}) is the modified Hamiltonian constraint which generalizes the usual Friedmann equation of GR. This equation constrains the 
possible values of $a,H,R,\dot R$ at some initial time. The explicit appearance of the scale factor $a$ will come from the contribution of the 
energy-momentum tensor of the matter. We call these values the {\it initial data}, although this 
does not mean that it is the data at or near the big bang. Clearly, the above equations reduce to the standard equations of GR for $f(R)=R$. 
Moreover, taking $f(R)= R-2\Lambda$ one recovers the equations of GR endowed with a cosmological constant.

Now, the expression for the Ricci computed directly from the metric is given by
\begin{equation}
\label{Rt}
R= 6\left(\dot H + 2 H^2 + \frac{k}{a^2}\right) \,\,\,.
\end{equation}
Note that by using Eqs.~(\ref{Friedmod}) and (\ref{accmod}) in Eq.~(\ref{Rt}) we obtain an identity $R\equiv R$, which shows the consistency 
of the equations. Thus, as emphasized before, this equation is redundant. However, one has the freedom of using the system of 
Eqs.~(\ref{traceRt}), (\ref{Friedmod}) and (\ref{Rt}) instead of Eqs.~(\ref{traceRt})$-$(\ref{accmod}) to solve for $(H,R)$, and Eq.~(\ref{Hubble}) to solve 
for $a$. In our case, we have in fact, used both sets of equations in order to check the consistency of our computer codes and also the accuracy of 
the numerical results at every ``time'' step (see below). We stress that Eq.~(\ref{Friedmod}) is only used to fix the initial data and then it is monitored to check 
that it is fulfilled at every integration step within the accuracy of the numerical algorithm (4th order Runge-Kutta scheme) when 
integrating the dynamical equations. At this regard, Motohashi {\it et al.}~\cite{Motohashi2010} had also used a similar technique to check the 
numerical accuracy.

As we mentioned before, the matter variables $T_{ab}$ obey their own dynamics 
which is provided by $\nabla_a T^{ab}=0$. We shall assume that  $T_{ab}$ is a mixture of 
three kinds of perfect fluids, $T_{ab}= \sum_{i=1}^3 T_{ab}^i$, $T_{ab}^i= (\rho_i + p_i)u_a u_b + g_{ab}\,p_i$, 
which correspond to baryons, radiation, and dark matter, respectively, in a epoch where they do not interact with each other except gravitationally. 
The energy-momentum tensor for each matter component is conserved separately, and the conservation equation $\nabla_a T^{ab}_i=0$ leads to the usual 
expression
\begin{equation}
\label{consT}
\dot \rho_i= -3 H \left(\rho_i + p_i\right)\,\,\,,
\end{equation}
where the total energy-density of matter is $\rho= \sum_{i=1}^3 \rho_i$. The above equation integrates straightforwardly (with $p_{\rm bar,DM}=0$ and 
$p_{\rm rad}= \rho_{\rm rad}/3$) as follows
\begin{equation}
\label{rhocosm}
\rho= \frac{\rho_{\rm bar}^0 + \rho_{\rm DM}^0}{(a/a_0)^3} + \frac{\rho_{\rm rad}^0}{(a/a_0)^4} \,\,\,,
\end{equation}
where the knotted quantities indicate their values today. 
Thus in this case the matter variables that appear in the field equations are given explicitly by $T^t_{\,\,t}= -\rho$ and 
$T= \sum_{i=1}^3 T_i= \sum_{i=1}^3 (3p_i-\rho_i)= 
-(\rho_{\rm bar}+ \rho_{\rm DM}) $ (since $T_{\rm rad}\equiv 0$, and $p_{\rm bar,DM}=0$), which in turn can be written 
in  terms of the scale factor according to Eq.~(\ref{rhocosm}). In this way, the differential equations will depend explicitly on $a(t)$. 

In the majority of the cosmological studies performed so far the equation for $\dot H$ is usually written in terms of $\ddot R$ which is clearly 
not suitable for an initial value problem (e.g. see Ref.~\cite{Dunsby2010} for a system of equations of this sort used for reconstructing $f(R)$ functions), 
unless one uses Eq.~(\ref{traceRt}) to replace such a term in favor or lower derivatives or alternatively 
if one uses a combination of several variables (e.g. see Sec. 4.1 of Ref.~\cite{deFelice2010}, and \cite{Amendola2007b}). 
Our approach is similar to that of Appleby \& Batty~\cite{Appleby2008} where they solve the set of Eqs. (\ref{traceRt}), (\ref{Hubble}) and (\ref{Rt}), but 
it is different from theirs in that they do not use Eq.~(\ref{Friedmod}) to fix the initial data and monitor the consistency of the numerical solutions 
(see Sec.~\ref{sec:numint}). In Sec.~\ref{sec:numint} we shall rewrite the system of equations using another ``time coordinate'' which is more suitable for the 
numerical integration. This is another important difference with respect to Ref.~\cite{Appleby2008} where the authors use $t$ itself as independent variable 
and furthermore they integrate backwards in time, which as we argue in Sec.~\ref{sec:numint} may have several inconveniences.


\subsection{The Equation of State in $f(R)$}
\label{sec:EOS}
Let us now study the specific application of the EMT's considered in Sec.~\ref{sec:EMT}. In GR with matter and dark energy components, 
the parameter $\ddot a$ is directly related to the total EOS $\omega_{\rm tot}= p_{\rm tot}/\rho_{\rm tot}$ 
[i.e. $\ddot a/a= -(\kappa \rho_{\rm tot}/6) (1+ 3\omega_{\rm tot})$], which in turn, determines if the universe is expanding in accelerating or decelerating way, 
provided $\rho_{\rm tot}>0$. Here $\rho_{\rm tot}= \rho + \rho_{\rm DE}$, and $p_{\rm tot}= p_{\rm rad} + p_{\rm DE}$. One can then define 
$\omega_{\rm DE}= p_{\rm DE}/\rho_{\rm DE}$, as the EOS for the dark-energy, which will coincide with $\omega_{\rm tot}$ when the matter 
contribution is negligible as compared to that of the dark-energy; this occurs in the $\Lambda CDM$ paradigm at future time.

We can in a similar way define an EOS for the component associated with $f(R)$ gravity. In order to achieve our goal we need then to define the 
energy-density $\rho_X$ and pressure $p_X$, associated with the geometric dark energy fluid. We shall obtain three expressions corresponding to Recipes I--IV. 
\bigskip

{\bf Recipe I:} First, we define the energy-density $\rho_X$ so that the modified Friedmann Eq.~(\ref{Friedmod}) reads 
(hereafter we asume $k=0$) 
\begin{equation}
\label{Hgen}
H^2=\frac{\kappa}{3}\left(\rule{0mm}{0.3cm} \rho +\rho_{X}\right) \,\,\,,
\end{equation}
where we remind the reader that $\rho$ includes all the contributions of ordinary matter (baryons and radiation) and dark matter as well. 
Second, we define its pressure $p_X$ so that Eq.~(\ref{accmod}) reads
\begin{equation}
\label{Hdotgen}
\dot{H}+H^2=-\frac{\kappa}{6}\left\{\rule{0mm}{0.4cm} \rho +\rho_{X}+3\left(p_{\rm rad}+ p_{X}\right) \right\} \,\,\,. 
\end{equation}

In this way, Eqs. (\ref{Friedmod}) and (\ref{accmod}) together with the above two equations lead to the following expressions:
\begin{equation}
\label{rhoX}
\rho_{X}=\frac{1}{\kappa f_{R}}\left\{\rule{0mm}{0.5cm} \frac{1}{2}\left( f_{R}R-f\right) -3f_{RR}H\dot{R} + 
\kappa \rho\left(1- f_{R}\right) \right\}\,,
\end{equation}

\begin{equation}
\label{pressX}
p_{X}=-\frac{1}{3\kappa f_{R}}\left\lbrace \frac{1}{2}\left(f_{R}R+f \right) + 3f_{RR}H\dot{R}-\kappa\left(\rho -3 p_{\rm rad} f_R \right)
\right\rbrace \,\,\,.
\end{equation}

The EOS $\omega_{X}$ of the $X-$fluid reads then
\begin{eqnarray}
\label{EOSX1a}
\omega_{X}= \frac{p_{X}}{\rho_{X}}\,.
\end{eqnarray}
The above equation can also be written in the following implicit form when using Eqs.~(\ref{Rt}), (\ref{Hgen}) and (\ref{Hdotgen}) 
\begin{equation}
\label{EOSX1b}
\omega_{X}= \frac{3H^2-3\kappa\,p_{\rm rad}-R}{3\left(3H^2-\kappa\rho\right)}\,\,\,.
\end{equation}
Clearly we have assumed here that $f(R)\neq R$~\footnote{Of course $f(R)= C={\rm const.}$ is also excluded as otherwise one is lead to 
consider a {\it vanilla gravity} described by $ C g_{ab}/2= -\kappa T_{ab}$. Moreover, this would lead to $f_{R}\equiv 0$ and $\rho_X$ and 
$p_X$ lead also to $0/0$.}, as otherwise one is led to $\omega_{X}=0/0$. Numerical examples of this EOS will be provided in Sec.~\ref{sec:numerics} 
for the three $f(R)$ models introduced in Sec.~\ref{sec:models} (c.f. Figs.~\ref{fig:wxallMir}--\ref{fig:wxallHu}).

One can recover Eqs.~(\ref{rhoX}) and (\ref{pressX}) directly from Eq.~(\ref{EMTX1a}) using 
$\rho_{X}:= u^a u^b T_{ab}^{X}$, with $u^a= (\partial/\partial t)^a)$,
 $p_{X}:= S^{X\,a}_{\,\,\,a}/3$, where $S_{ab}^X$ is the 3-energy-momentum tensor obtained from $T_{ab}^X$ and which is defined on the orthogonal hypersurfaces to $u^a$
\footnote{$S^{X\,a}_{\,\,\,a}\equiv T^{X\,i}_{\,\,\,i}$ is the spatial trace of $T_{ab}^X$.}, and by replacing $\ddot R$ (which will appear in $\rho_X$ 
via the term $\nabla_t \nabla_t R$) in favor of lower order derivatives using Eq.~(\ref{traceRt}).
 
As we have emphasized, $T_{ab}^{X}$ is conserved $\nabla^a T_{ab}^{X}=0$. 
Indeed, since the only energy-momentum tensor compatible with the hypothesis of homogeneity and isotropy is that of an effective 
perfect fluid, then {\it a fortiori} $\rho_{X}$ and $p_{X}$ must obey an 
equation similar to Eq.~(\ref{consT}). This statement can be checked explicitly by using Eqs.~(\ref{rhoX}) and (\ref{pressX}) in Eq.~(\ref{consT}), and then 
appealing to the field equations. 

The other general considerations discussed in Recipe I shall apply to this particular case. For instance, 
although the matter terms appear in Eqs. (\ref{rhoX}) and (\ref{pressX}), 
$\rho_{X}=0= p_{X}$ for the GR case $f(R)=R$, and moreover, $\rho_{X}= \Lambda/\kappa=-p_X$, $\omega_X=-1$, for $f(R)= R-2\Lambda$. 
This shows that these quantities so defined are sensible because it is only when the theory differs from GR that they do not vanish in general, and 
when $\Lambda$ is included, they reduce to the expected values. 
Of course, we do not intend to add a cosmological constant in the $f(R)$ models, the only purpose of these remarks is 
to illustrate some of the properties of these definitions.  
Notice that these expressions do not correspond to the total energy-density and pressure of the effective ``matter'', {\it i.e.} the sum of ordinary matter, 
dark matter and $X$--matter, but only to the $X$--fluid. The total equation of state is given by
\begin{equation}
\label{wtot}
\omega_{\rm tot}:= \frac{p_{\rm tot}}{\rho_{\rm tot}}\,\,\,,
\end{equation}
where $\rho_{\rm tot}= \rho_X + \rho$ [c.f. Eq.~(\ref{Hgen})]~\footnote{Under this definition $\Omega:= \kappa \rho_{\rm tot}/3H^2=1$ for $k=0$ 
[c.f. Eq.~(\ref{Omegatot})].} and $p_{\rm tot}= p_X + p_{\rm rad}$ [c.f. Eq.~(\ref{Hdotgen})]. The total EOS is depicted in Fig.~\ref{fig:wtot}. 
These quantities can also be computed directly from $T_{ab}^{\rm tot}$ as $\rho_{\rm tot}= u^a u^b T_{ab}^{\rm tot}$ and $p_{\rm tot}:= T^{{\rm tot}\,i}_{\,\,\,i}/3$.  

As part of Recipe I, one can write $\rho_{X}$ and $p_{X}$ in terms of ``purely geometric'' quantities where the 
matter terms do not appear explicitly. This is in fact achieved by considering the equivalent definition Eq.~(\ref{EMTX1b}). 
Let us see explicitly how can we arrive, for instance, to $\rho_X$ directly from the field equation, which is easier than using 
Eq.~(\ref{EMTX1b}). Consider Eq.~(\ref{Friedmod}), with $k=0$ for simplicity, and again assume a perfect fluid. Then we have 
\begin{equation}
\label{cero}
0 = -3H^2 f_R + \frac{1}{2}\left( f_R R- f\right) - 3f_{RR} H \dot R + \kappa \rho\,\,\,.
\end{equation}
Adding $3H^2$ to both sides we then write
\begin{equation}
H^2= \frac{1}{3}\left[ 3H^2(1-f_R) + \frac{1}{2}\left( f_R R- f\right) - 3f_{RR} H \dot R\right] + \frac{\kappa}{3} \rho \,\,\,.
\end{equation}
Therefore, this equation reads like Eq.~(\ref{Hgen}) if one defines the term in brackets as follows
\begin{equation}
\label{rhoXalt}
\kappa \rho_X= 3H^2(1-f_R) + \frac{1}{2}\left( f_R R- f\right) - 3f_{RR} H \dot R \,\,\,.
\end{equation}
In this way, the matter term does not appear explicitly in $\rho_X$, but clearly (\ref{rhoXalt}) and (\ref{rhoX}) are 
one and the same thing. We can recover the previous equation if one uses $\kappa \rho= 3H^2 -\kappa \rho_X$ 
in Eq.~(\ref{rhoX}) and then solves for $\rho_X$. One can rewrite $p_X$ in a similar fashion (c.f. Ref.~\cite{Miranda2009}). 

Miranda {\it et al.}~\cite{Miranda2009}, 
de Felice \& Tsujikawa~\cite{deFelice2010},  Motohashi {\it et al.}~\cite{Motohashi2010,Motohashi2011a,Motohashi2011b}, 
and Bamba {\it et al.}~\cite{Bamba} have used a definition for the EOS which coincides with Recipe I. In the case of HS1~\cite{Hu2007}, their definition 
coincides exactly with our expression Eq.~(\ref{EOSX1b}) when the radiation contribution is completely neglected\footnote{In HS1 the authors use completely 
different variables.}, which is justified only in the matter dominated epoch. It is to be notice, that some authors have preferred to keep the terms like 
$\ddot R$ and $\ddot a$ in their definition, instead of writing them using lower order derivatives as we do.

Equations~(\ref{rhoX})--(\ref{EOSX1a}) or even Eq.~(\ref{EOSX1b}), are rather simple and easier to calculate under our approach. They are evaluated easily 
at every integration step when solving numerically our system of equations (see Sec.~\ref{sec:numint}).
\bigskip

{\bf Recipe II:} In order to obtain $\rho^{II}_X$ associated with Recipe II we can use Eq.~(\ref{EMTX2a}) or (\ref{EMTX2b}). However, for our purposes 
Eq.~(\ref{EMTX2b}) is more convenient. We obtain then 
\begin{equation}
\label{rhoX2}
\rho^{II}_X=\frac{A}{\kappa f_{R}}\left\{\rule{0mm}{0.5cm} \frac{1}{2}\left( f_{R}R-f\right) -3f_{RR}H\dot{R} + 
\kappa \rho\left(1- \frac{f_{R}}{A}\right) \right\}\,,
\end{equation}
\begin{equation}
\label{pressX2}
p^{II}_X= -\frac{A}{3\kappa f_{R}}\left\lbrace \frac{1}{2}\left(f_{R}R+f \right) + 3f_{RR}H\dot{R}-\kappa\left(\rho -3 p_{\rm rad}\frac{f_R}{A}\right)  
\right\rbrace \,\,\,.
\end{equation}
Equation (\ref{rhoX2}) can be recovered easily from Eq.~(\ref{Friedmod}), following almost the same steps as in Recipe I, except that now we define $\rho^{II}_X$ from
\footnote{Eq.~(\ref{rhoX2}) has the alternative expression
\begin{equation}
\label{rhoXAGPT}
\kappa \rho^{II}_X= 3H^2(A-f_R) + \frac{1}{2}\left( f_R R- f\right) - 3f_{RR} H \dot R \,\,\,.
\end{equation}
The interested reader can consult Ref.~\cite{Gannouji2009} for the ``purely geometric'' expression for $p^{II}_X$.}
\begin{equation}
\label{HgenII}
AH^2=\frac{\kappa}{3}\left(\rule{0mm}{0.3cm} \rho + \rho^{II}_X\right) \,\,\,.
\end{equation}
Therefore the EOS reads 
\begin{equation}
\omega^{II}_X=\frac{p^{II}_X}{\rho^{II}_X}  \,.
\end{equation}
The quantities of Recipe I are recovered for $A=1$. In Refs.~\cite{Tsujikawa2007,Amendola2007b,Amendola2008,Gannouji2009} Recipe II 
was used with $A=f_R^0$ ($F_0$ in their notation; see Ref.~\cite{Gannouji2009} for a discussion concerning the introduction of this constant $f_R^0$).

In general $f_R^0 \neq 1$ although $f_R^0 \approx 1$ (c.f. Ref.~\cite{Hu2007}) 
\footnote{Since this dimensionless quantity appears as $G_{\rm eff}= G_0/f_R$ and at present time $G_{\rm eff}^0\approx G_0$, 
thus one expects $f_R^0 \approx 1$.}. As shown by Hu \& Sawicky~\cite{Hu2007}, one has to take (at least in their model) $f_R^0 \approx 1$ to 
avoid large deviations in the EOS relative to the cosmological constant value $\omega_\Lambda=-1$. So if one takes $f_R^0 \neq 1$, $\omega_X^{II}$ and 
$\omega_X$ are not equivalent. As a matter of fact, $\omega_X^{II}$ can diverge at some redshift depending on the $f(R)$ model. For instance, in 
the Starobinsky and MJW model this is exactly what happens (c.f. Figure~\ref{fig:wx2-St-Mir}), 
because $\rho^{II}_X$ becomes zero (c.f. Figs.~\ref{fig:rhoXMir} and \ref{fig:rhoXSt}). In the Hu--Sawicky model we did not encounter 
that divergence in the range of redshifts we explored (c.f. Fig.~\ref{fig:wxallHu}). Due to this pathological behavior, this definition for the EOS 
makes it rather unsuitable for comparing with observations.
\bigskip

{\bf Recipe III:} In this case $\rho^{III}_X$  and $p^{III}_X$ are obtained from Eq.~(\ref{EMTX3}). In particular, $\rho^{III}_X$ can be 
read off directly from Eq.~(\ref{Friedmod}), by defining $H^2= \kappa(\rho^{III}_X + \rho)/(3f_R)$, and get
\begin{equation}
\label{rhoXIII}
\rho^{III}_X = \frac{1}{\kappa}\left[\frac{1}{2}\left(f_{R}R-f\right) -3f_{RR}H\dot{R}\right]\,.
\end{equation}
The pressure and the EOS read respectively
\begin{equation}
\label{pressXIII}
p^{III}_X= -\frac{1}{3\kappa}\left[3f_{RR}H\dot{R} + \frac{1}{2}\left(f_{R}R + f\right) + \kappa T\right]\,,
\end{equation}
and
\begin{equation}
\omega^{III}_X=\frac{p^{III}_X}{\rho^{III}_X}\,,
\end{equation}
where $T= 3p_{\rm rad} - \rho$, as before. This EOS coincides with $\omega_X$ only in vacuum. But even in 
vacuum the density and pressure does not satisfy a conservation equation like~(\ref{consT}), as it was stressed in Sec.~\ref{sec:EMT}.
\bigskip

{\bf Recipe IV:} In this case $\rho^{IV}_X$ and $p^{IV}_X$ arise from Eq.~(\ref{EMTX4}) or from the definition 
$T_{ab}^{IV\,,\,X} := f_R^{-1} T_{ab}^{III\,,\,X}$ as given in Sec.~\ref{sec:EMT}, which yields
\begin{eqnarray}
\label{rhoXIV}
\rho^{IV}_X = \frac{\rho^{III}_X}{f_{R}}\,\,,
\end{eqnarray}
\begin{eqnarray}
\label{pressXIV}
p^{IV}_X= \frac{p^{III}_X}{f_{R}}\,\,.
\end{eqnarray}
The EOS is thus
\begin{equation}
\omega^{IV}_X=\frac{p^{IV}_X}{\rho^{IV}_X}\equiv \omega^{III}_X\,.
\end{equation}
Alternatively, $\rho^{IV}_X$ can be read off directly from Eq.~(\ref{Friedmod}), by defining $H^2= \kappa\left(\rho^{IV}_X + \rho/f_R\right)/3$.

Notice that the Recipe IV quantities coincide with $\rho_{X}$, $p_{X}$ and $\omega_{X}$ only in vacuum. Nonetheless, in the non vacuum case 
$\rho^{IV}_X$ and $p^{IV}_X$ have the unaesthetic feature of not satisfying a conservation equation like ~(\ref{consT}), but rather present an extra source 
term that depends on the total matter density
\footnote{The explicit form of the source term can be appreciated in Eq.~(108) of Ref.~\cite{Capozziello2008a}, Eq.~(8) of Ref.~\cite{Capozziello2008b} 
and Eq.~(11.3) of Ref.~\cite{Bamba2012}.}.
Quantities associated with Recipe IV have been considered by Sotiriou \& Faraoni~\cite{Sotiriou2010}
\footnote{In Sotiriou \& Faraoni~\cite{Sotiriou2010} it was defined an EMT denoted as $\,T_{\mu\nu}^{(eff)}$ which corresponds to our 
$T_{ab}^{III\,,\,X}$. Incidentally, those authors define also $\rho_{eff}$ and $p_{eff}$ which correspond to our $\rho^{IV}_X$ and 
$p^{IV}_X$, rather to $\rho^{III}_X$ and $p^{III}_X$. This can be appreciated in Ref.~\cite{Sotiriou2010} as the quantities $\rho_{eff}$ and $p_{eff}$ 
do contain the the alluded factor $f_R^{-1}$, which is missing in their $\,T_{\mu\nu}^{(eff)}$. It seems that those authors defined $\rho_{eff}$ and $p_{eff}$ 
using the cosmological field equations instead of using $\,T_{\mu\nu}^{(eff)}$ directly. In any case, this apparently lack of consistency in their notation has 
no effect on the EOS as Recipes III and IV give rise to the same EOS. However, the reader is urged to have in mind such nuances 
in order to avoid any confusion between the current paper and theirs. Still, none of the $X$--fluid quantities associated with $T_{ab}^{III\,,\,X}$ or 
$T_{ab}^{IV\,,\,X}$ will satisfy a conservation equation like Eq.~(\ref{consT}).},
Capozziello {\it et al.}~\cite{Capozziello2008a, Capozziello2008b} 
\footnote{A typographical error (which seems to be systematic) is found in Capozziello {\it et al.} Refs.~\cite{Capozziello2008a, Capozziello2008b} 
where a sign `$-$' seems to be missing in front of the term $\frac{1}{2}\left( f- f_{R}R\right)$ when they define $\rho_{\rm curv}= 
\frac{1}{\kappa f_{R}}\left\lbrace \frac{1}{2}\left( f- f_{R}R\right) - 3f_{RR}H\dot{R} \right\rbrace$ [c.f. Eq.~(\ref{rhoXIII}) and Eq.~(\ref{Friedmod}) 
in vacuum $\rho=0=-T^t_{\,\,t}$ ]. This is not a global sign, so their definition for $\rho_{\rm curv}$ seems to be inconsistent with their modified 
Friedmann equation which (modulo notation) should be identical to ours. 
Furthermore, taking $f(R)= R-2\Lambda$ 
yields $\rho_{\rm curv}= - \Lambda/\kappa$ which has the opposite sign that one should usually consider for $\Lambda >0$. 
The expression for $\rho_{\rm curv}$  is of course invariant under $f\rightarrow -f$, 
thus the difference cannot be explained by a sign convention in the gravitational Lagrangian.},
and Bamba {\it et al.}~\cite{Bamba2012}, although their expression for the pressure is written in a different way (a term $\ddot R$ appears explicitly there). 

Despite the unpleasant features associated with $\omega^{III}_X=\omega^{IV}_X$, they can behave similarly to $\omega_X$ for actual cosmological scenarios 
(c.f. Figs.~\ref{fig:wxallMir}--\ref{fig:wxallSt}).

In summary, there exist at least three inequivalent definitions $\omega_{X}$, $\omega_{X}^{II}$, $\omega^{III}_X$ that we have identified in the 
literature as arising from different sorts of EMT representing the geometric dark energy component. 
We consider that $\omega_{X}$ has the most appealing properties, while EOS $\omega_{X}^{II}$ may have serious deficiencies as it can diverge 
at some redshifts. On the other hand, $\omega_{X}^{III}=\omega_{X}^{IV}$ have the unaesthetic feature of being 
related to a non conserved EMT. But even if we dismiss this feature, there are quantitative differences between 
$\omega_{X}$ and $\omega_{X}^{III}$ (or $\omega_{X}^{IV}$) and although the numerical discrepancies between them maybe considered unimportant at present time for 
the viable cosmological models, they may, however, become important if the EOS is measured with high precision in the future. In this latter instance, one should 
bare in mind which definition is being used to make comparisons with the values that are inferred from observations. We cannot emphasize more this last comment.

In the following sections we describe the numerical strategy we use to integrate the system of equations (independently of the specific $f(R)$ model considered) 
and then analyze the cosmological results for the three specific models described in Sec.~\ref{sec:models}.


\subsection{Numerical integration}
\label{sec:numint}

The system of Eqs.~(\ref{traceRt}), (\ref{accmod}) and (\ref{Hubble})
as well as the alternative system  (\ref{traceRt}), (\ref{Hubble}) and (\ref{Rt}) 
have the form $dy^i/dt= {\cal F}^i(y^i)$ where $y^i= (a,H,R,\Pi)$ and $\Pi:= \dot R$. Therefore they can be solved easily with a fourth order 
Runge-Kutta algorithm. Now, it is well known in cosmology that $t$ is not the best independent variable to perform 
the cosmic evolution because usually several scalars blow up very fast as $a\rightarrow 0$. It turns better to use the following 
parameter to integrate the differential equations~\footnote{Clearly, $\alpha$ is a good ``time coordinate'' provided $a(t)$ is a monotonic function, 
otherwise the relationship between 
$t$ and $\alpha$ is not in one-to-one correspondence, making the mapping $t(\alpha)$ ill defined.}
\begin{equation}
\label{alpha}
\alpha = {\rm ln}(a/a_0) \,\,\,,
\end{equation}
which maps the big bang $a\rightarrow 0$ to $\alpha\rightarrow -\infty$. So when integrating the equations with respect to this variable 
one is always far from the big bang in the $\alpha$ domain, but one can be very close to it in the $t$ domain, in the sense that $a/a_0\ll 1$. 
Equations (\ref{traceRt}), (\ref{Friedmod}), (\ref{accmod}), and (\ref{Rt}) read as follows with respect to $\alpha$:
\begin{equation}
\label{R-numerical}
R''=-R' \left(1+\frac{R}{6H^{2}}\right)-\frac{1}{3f_{RR}H^{2}}\left[ 
3f_{RRR}H^{2}R'^{2}+2f-f_{R}R+\kappa T \right]\,\,,
\end{equation}

\begin{equation}
\label{dHnumerical}
H'=-2H+\frac{R}{6H}\,\,,
\end{equation}

\begin{equation}
\label{Friedmod-numerical}
H^2 + \frac{1}{f_R}\left[ f_{RR} H^{2} R' -\frac{1}{6}\left( f_R R- f\right)\right]= \frac{-\kappa T^t_{\,\,t}}{3f_R}\,\,,
\end{equation}

\begin{equation}
\label{accmod-numerical}
H' = -H + \frac{1}{f_{R}H}\left(f_{RR}  H^{2} R' + \frac{f}{6} + \frac{\kappa T^t_{\,\,t}}{3} \right) \,\,.
\end{equation}

As we have mentioned, we can use Eq.~(\ref{dHnumerical}) or Eq.~(\ref{accmod-numerical}) to solve 
for $H$. Nevertheless, we have used both equations to verify the consistency of our numerical code and together with 
the verification of the modified Hamiltonian constraint at every integration step, prove the soundness of our results.

From (\ref{Hubble}), and (\ref{alpha}) one obtains
\begin{eqnarray}
\label{invHubble}
\frac{dt}{d\alpha}= 1/H\,\,\,,
\end{eqnarray}

which is used to extract the age of the Universe when inverting the numerical solution $t(\alpha)$ so as to obtain 
$a(t)$ (the age is ``read off'' from its graph). The initial condition needed to solve this 
equation simply sets some irrelevant initial time which can have any arbitrary value.  
The relevant chronological quantities are the differences between any given time and the initial time.

Different system of equations using different variables have been analyzed in the past in order to perform the integration 
(e.g. see \cite{Hu2007,Evans2008,Amendola2008,Amendola2007b,deFelice2010}). In particular, the system employed by HS1~\cite{Hu2007} has become very 
popular in recent years. 

We stress that the numerical integration must be performed forward in $\alpha$ because there are solutions in $f(R)$ that behave like an attractors 
as one evolves towards the future. Integrating in the opposite direction can then easily lead to cosmological solutions that do not satisfy (retrodict) the physical 
conditions they should have in the past (e.g. those dictated by CBR or nucleosynthesis). We think that this is precisely the problem that was encountered 
in Ref.~\cite{Appleby2008}, where some kind of ``singularity'' was found due to runaway solutions which corresponded to ``inadequate'' initial data. Such data, 
that we can call $d_f^c$ and which is fixed in the ``future'', might be slightly different from the data that is predicted when integrating 
forward in time using $d_p^c$ as initial data fixed in the past, but this difference is large enough not only to prevent the system from retrodicting 
the initial data $d_p^c$, but also to make it incapable of arriving to the same past epoch associated with 
$d_p^c$. One would then require a remarkable precision on $d_f^c$ to retrodict $d_p^c$; such 
are the features of dynamical systems that have attractors in one direction of the evolution parameter. It is beyond the scope of the present paper to analyze 
such dynamical properties of the system. 

The numerical strategy we used to integrate the equations is as follows : 
We start the integration at some initial $z$ in the past ($z_c$), where $z= a_0/a -1= e^{-\alpha}-1$, given 
initial conditions for $\rho_c^{\rm matt}$, $\rho^{\rm rad}_c$, $R_c$, $R'_c$, $H_c$.  
From Eqs.~(\ref{Friedmod-numerical}) and (\ref{EOSX1b}) evaluated at the initial $z_c$, one solves algebraically 
for $R^\prime_c$ and $R_c$, respectively, in terms of the remaining variables. In this way, both $R_c$ 
and $R'_c$ are fixed given the initial values $H_c$, $\rho_c^{\rm matt}$, $\rho^{\rm rad}_c$ and $\omega_X^c$. We can fix 
$\rho_c^{\rm matt}$ and $\rho^{\rm rad}_c$ using the $\Lambda CDM$ as a guide to chose the abundances of baryons, 
DM and radiation at that $z_c$, and take $\omega_X^c \approx -1$. This value for $\omega_X^c$ 
might seem rather {\it ad-hoc} at first sight, however, since one expects that for ``high'' $z$ the curvature $R$ is also sufficiently high for the $f(R)$ models to 
behave as GR plus $\Lambda_{\rm eff}$ (c.f. Fig.\ref{fig:F(R)'s}) then one can anticipate that 
$\omega_X^c \approx -1$ is not a bad approximation. Moreover, we have adopted such value 
in order to compare our results with other authors who have taken similar values for the EOS at 
the same $z_c$. In any case, the value for $\omega_X^c$ is not a fundamental issue 
and initial conditions can be fixed as one pleases provided the modified Hamiltonian 
constraint is satisfied and that the resulting cosmological model is consistent with 
observations within the error bars. So, 
what we have used as main criteria to asses if of our initial data 
is adequate is that our results at present time be consistent with what is measured from observations. 
Finally, the remaining quantity to be fixed for the rest of the initial data to be 
determined is $H_c$. The $\Lambda CDM$ model of GR provides also a clue of the expansion rate 
at a given $z$, and in order to fix it more accurately we used a shooting-like method 
for $H_c$ that allowed to recover the 
actual abundances for each model at present time when integrating the system 
(\ref{R-numerical}), (\ref{dHnumerical}) and (\ref{accmod-numerical}). As we have stressed, 
Eq.~(\ref{Friedmod}) serves only as a constraint for the initial data and also 
to monitor the accuracy of 
the numerical results at every time step. This equation is extremely sensitive 
to any typing mistake in the computer code as well as to any 
inconsistency when fixing the initial data. A final comment is in order concerning the 
different recipes for the EOS. Since we use recipe I given by Eq.~(\ref{EOSX1b}) to help us fixing 
the initial data, we do not expect that the initial (and in general any) values for the other EOS, $\omega_X^{II}$, and $\omega_X^{III}$ 
will be the same as $\omega_X$. Alternatively, we could have opted to fix 
the initial data so that the three EOS initially behave the most similarly, knowing that proceeding in this way could ruin 
the prediction for the present values of the observables. This method can certainly be followed, but we do not pursue it here. The important point to keep in mind 
is that our resulting cosmological models are viable given a self-consistent initial data.

It is important to point out some differences about the fixing of initial data with respect to some recent works that have appeared in the literature. 
For instance, several authors~\cite{Poplawski2006,Capozziello2008b,Cardone2012} have proposed to use the current values of the deceleration parameter and the {\it jerk}, 
respectively related to $\ddot a_0$ and $\dddot a_0$, to fix the initial conditions for $R_0$ and $\dot R_0$. For instance, such 
quantities can be written respectively as \footnote{Notice that some authors define the jerk with the opposite sign~\cite{Benitez2012}.}
\begin{eqnarray}
\label{dec}
q &:=& -\frac{\ddot a}{aH^2}= -\frac{H^{2}+\dot{H}}{H^{2}} =  1- \frac{R}{6H^2}\,\,,\\
\label{jerk}
j &:=& \frac{\dddot a}{aH^3}= \frac{\dot{R}}{6H^3} -\frac{\dot H}{H^2} + 1= \frac{\dot{R}}{6H^3} + q + 2 \,\,.
\end{eqnarray}
This would be a nice strategy if one could easily integrate from 
present to past, nevertheless, as we have emphasized, one easily goes into inadequate solutions by proceeding that way. Following our integration strategy, 
the deceleration and jerk parameters are rather predicted from our initial conditions imposed in the past. In the next section we provide today's 
values of these parameters for each of the three $f(R)$ models analyzed here.
 

\section{Numerical results and cosmic viability}
\label{sec:numerics}

\begin{figure}[ht]
\includegraphics[width=9cm]{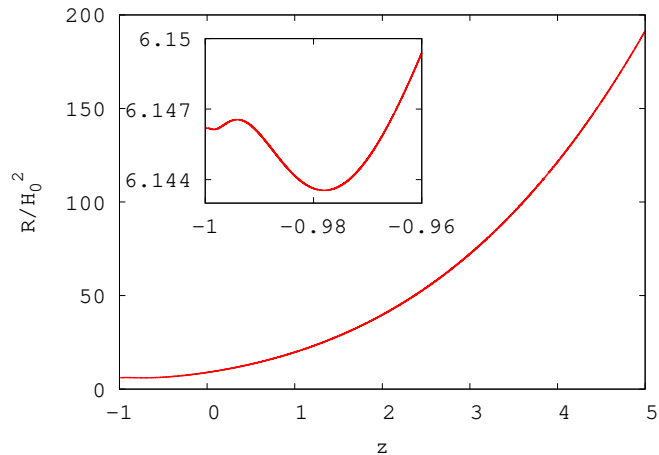}
\caption{Ricci scalar for the MJW model as a function of the light shift $z$.}
\label{fig:RMir}
\end{figure}

\begin{figure}
\includegraphics[width=9cm]{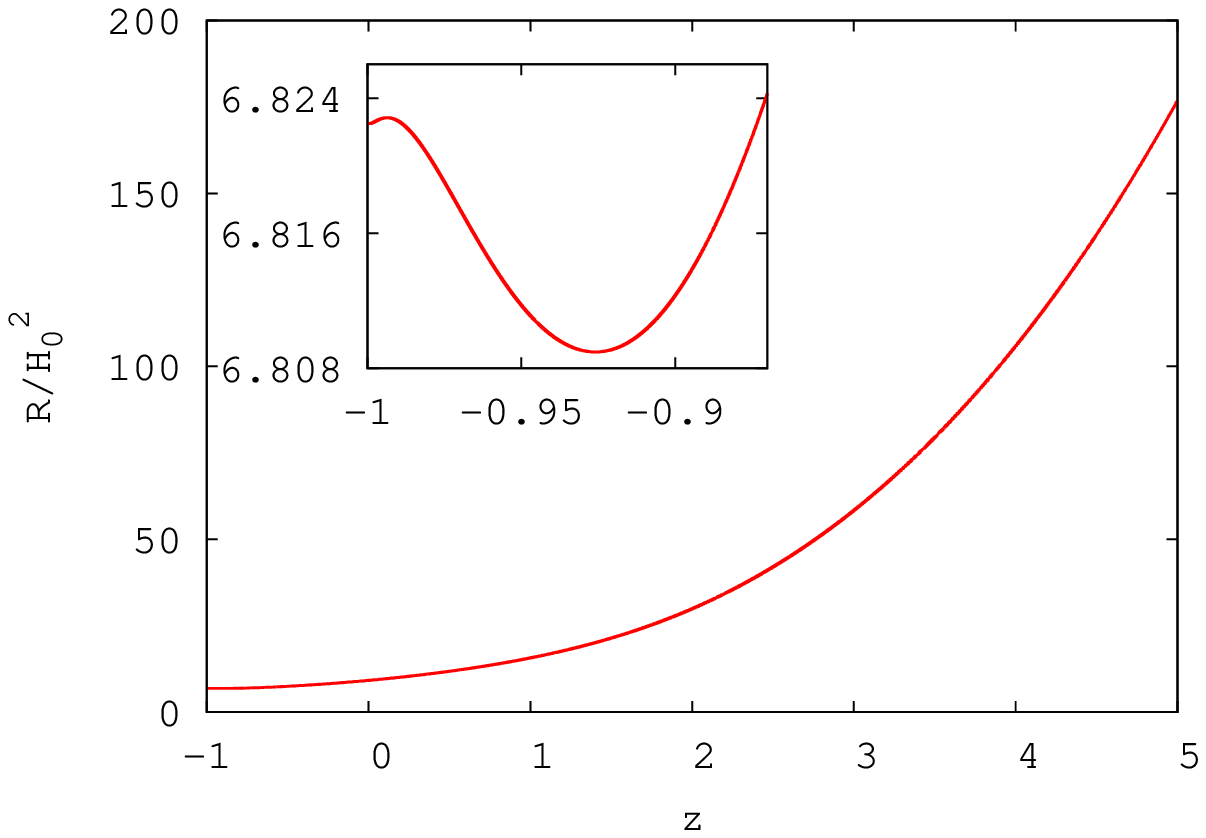}
\caption{Same as Fig.~\ref{fig:RMir} for the Starobinsky model.}
\label{fig:RSt}
\end{figure}

\begin{figure}[ht]
\includegraphics[width=9cm]{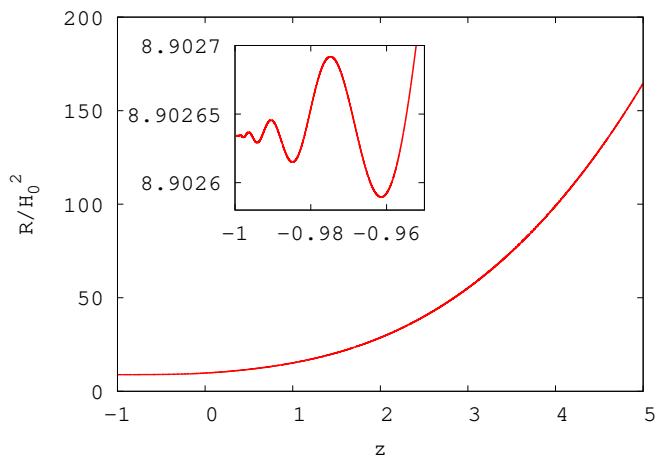}
\caption{Same as Fig.~\ref{fig:RMir} for the Hu--Sawicky model.}
\label{fig:RHu}
\end{figure}

In this section we present the numerical results for the three $f(R)$ models introduced in Sec.~III. Figures~\ref{fig:RMir}--\ref{fig:RHu} 
depict the Ricci scalar $R$ as a function of the light-shift $z=a_0/a-1$, where $z=0$ corresponds to the present time. 
For the three models, $R$ approaches the de Sitter minimum of the potentials 
plotted in Figures~\ref{VMiranda}--\ref{VHu}. The expansion $H$ damps $R$, and near the minimum, it oscillates. The oscillations can be easily 
explained by a linear perturbation around the de Sitter minimum where the matter contribution is negligible as it dilutes rapidly with the expansion.

\begin{figure}[ht]
\includegraphics[width=9cm]{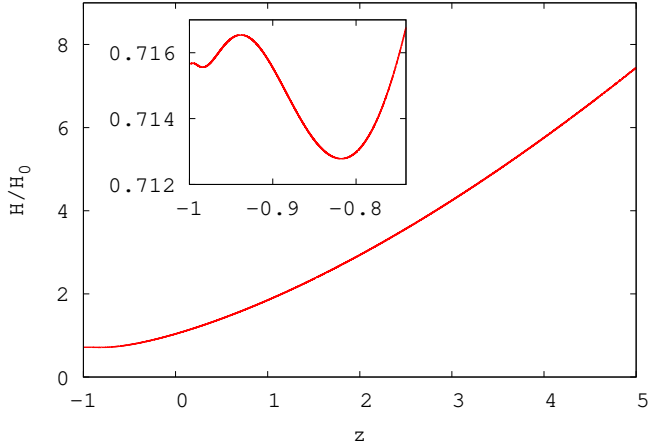}
\caption{Hubble expansion for the MJW model as a function of the light shift $z$.}
\label{fig:HMir}
\end{figure}

\begin{figure}[ht]
\includegraphics[width=9cm]{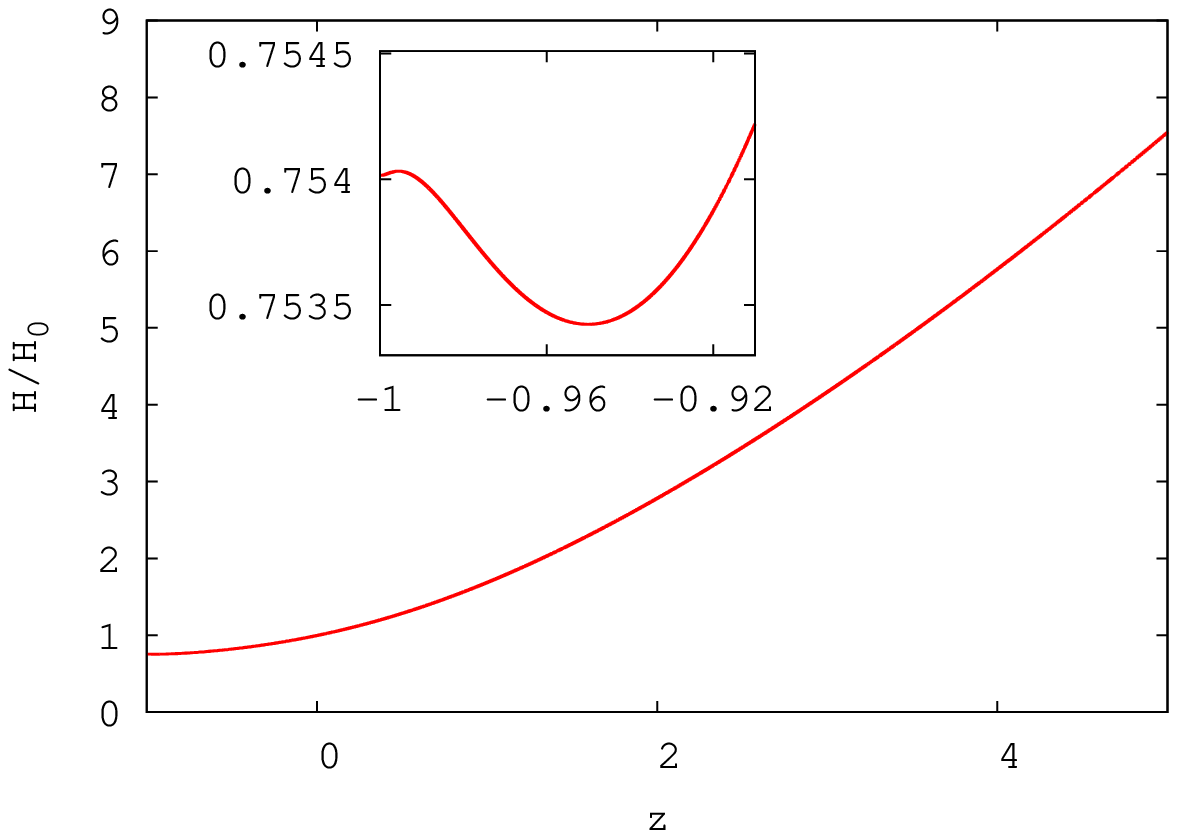}
\caption{Same as Fig.~\ref{fig:HMir} for the Starobinsky model.}
\label{fig:HSt}
\end{figure}

\begin{figure}[ht]
\includegraphics[width=9cm]{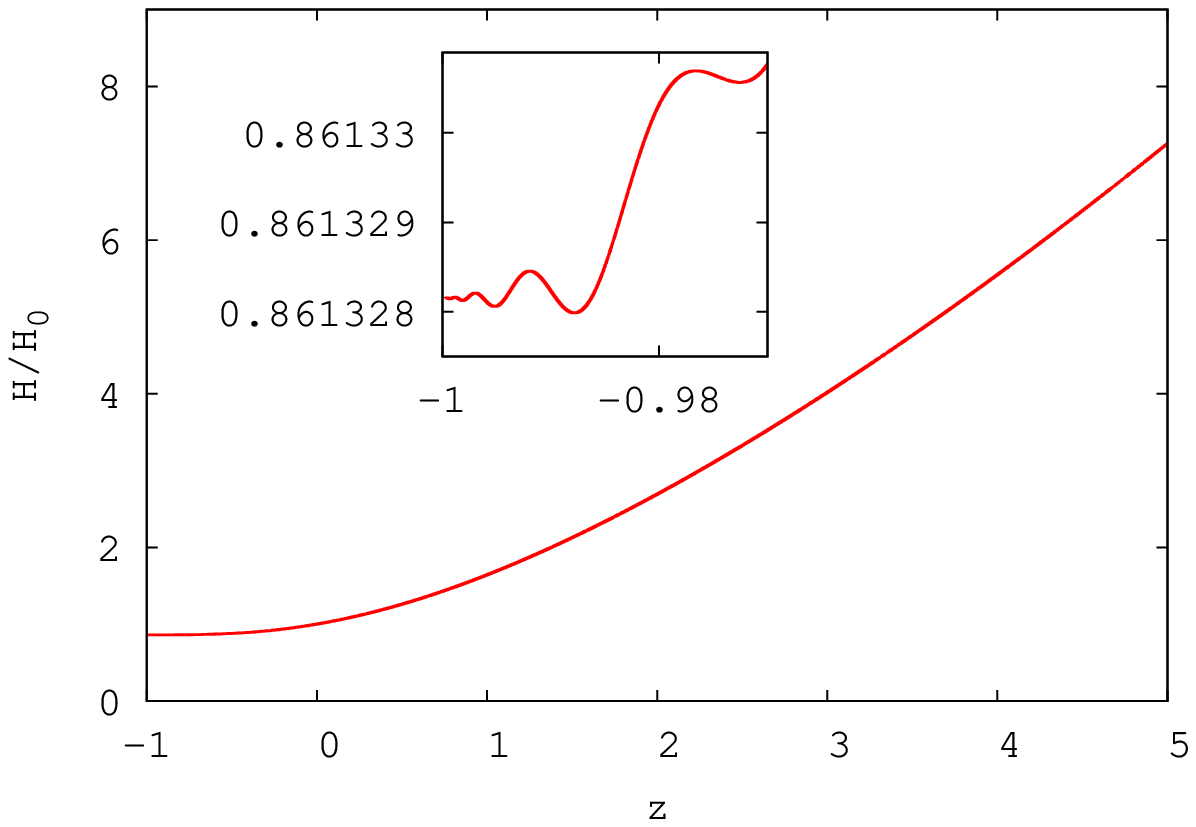}
\caption{Same as Fig.~\ref{fig:HMir} for the Hu--Sawicky model.}
\label{fig:HHu}
\end{figure}

Figures~\ref{fig:HMir}--\ref{fig:HHu} show the expansion rate $H$. Like in Refs.~\cite{Bamba}, we see that the Hubble expansion oscillates at late times for 
the three models.

We introduce now the dimensionless densities of the different ``species'' as follows:
\begin{equation}
\label{Omegatot}
\Omega:=  \Omega_{\rm rad} + \Omega_{\rm bar}+ \Omega_{\rm DM} + \Omega_X= 1 \,\,\,,
\end{equation}
where $\Omega_i= \kappa \rho_i/(3H^2)$~\footnote{Some authors \cite{Tsujikawa2007,Amendola2008,Tsujikawa2008,Gannouji2009} define 
$\Omega_i= \kappa \rho_i/(3 f_R H^2)$. This difference must be bare in mind when comparing results.}. Here $\rho_X$ is taken as in 
Recipe I.

\begin{figure}[ht]
\includegraphics[width=9cm]{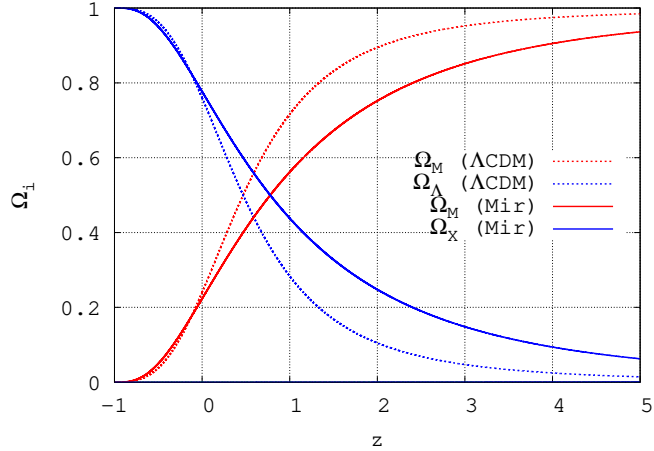}
\caption{Densities for the MJW model as a function of the light shift $z$. For reference the densities of the $\Lambda CDM$ model are also plotted. 
Radiation has been taken into account but cannot even be noticed during this epoch. Here we assumed $\,^{\Lambda CDM}\,\!\Omega_{\rm bar+DM}^0\approx 0.24$,  
$\,^{\Lambda CDM}\,\!\Omega_{\rm rad}^0 \approx 4.1 \times 10^{-5}$ and $\Omega_{\Lambda}^0\approx 0.76$ for the $\Lambda CDM$ model of GR. The predicted values 
of the corresponding densities today for the $f(R)$ model are slightly different as they are predicted given the initial conditions in the past.}
\label{fig:Ome-Mir}
\end{figure}

\begin{figure}[ht]
\includegraphics[width=9cm]{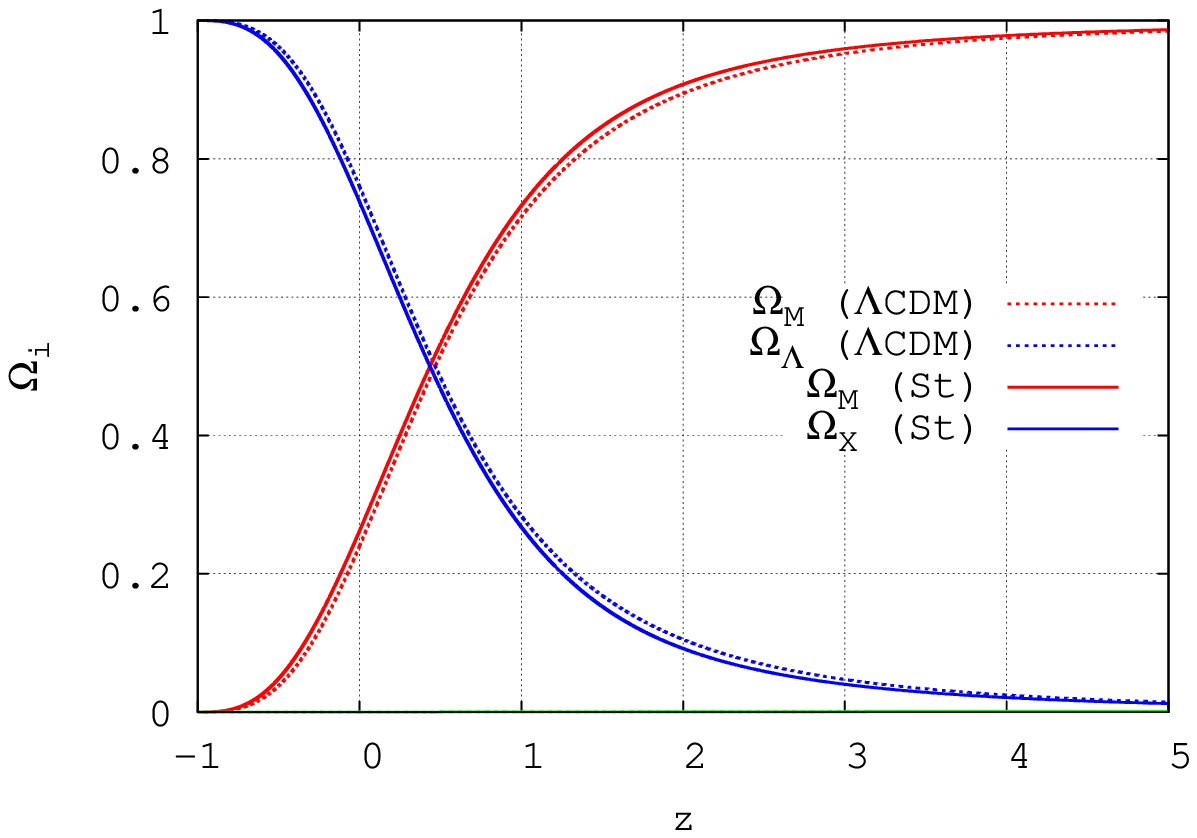}
\caption{Same as Fig.~\ref{fig:Ome-Mir} for the Starobinsky model.}
\label{fig:Ome-St}
\end{figure}

\begin{figure}[ht]
\includegraphics[width=9cm]{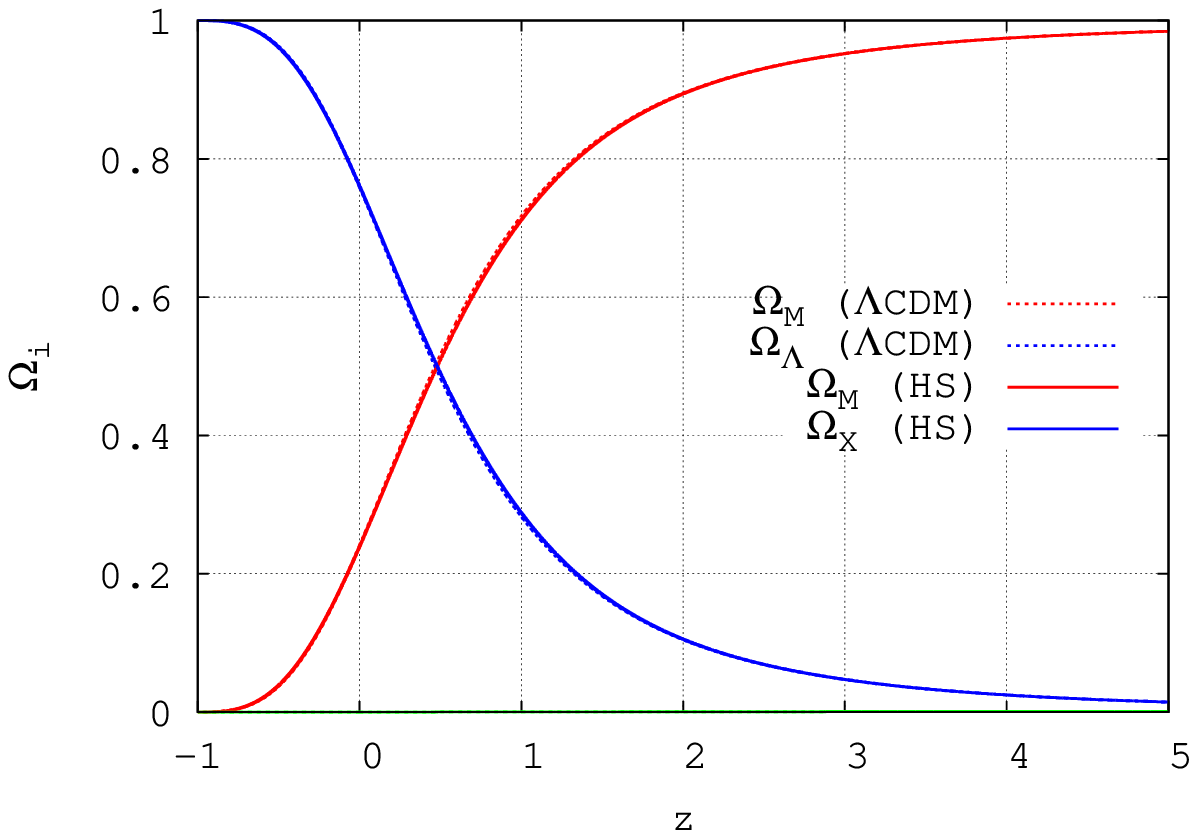}
\caption{Same as Fig.~\ref{fig:Ome-Mir} for the Hu--Sawicky model.}
\label{fig:Ome-Hu}
\end{figure}

Figures \ref{fig:Ome-Mir}--\ref{fig:Ome-Hu} depict the relative abundances $\Omega_i$ as a function of light-shift $z$. For reference, 
the corresponding abundances of the $\Lambda CDM$ model are also plotted. 
The abundances in the  $\Lambda CDM$ model have the following analytical expression in terms of their values today (knotted quantities) and the scale factor 
${\bar a}=a/a_0$: $\Omega_{i}^{\Lambda CDM}= \,^{\Lambda CDM}\,\!\Omega_i^0 {\bar a}^I\left[\left(\,^{\Lambda CDM}\,\!\Omega_{\rm bar}^0 
+ \,^{\Lambda CDM}\,\!\Omega_{\rm DM}^0\right){\bar a}^{-3} 
+ \,^{\Lambda CDM}\,\!\Omega_{\rm rad}^0{\bar a}^{-4} + \Omega_{\Lambda}^0\right]^{-1}$, where the subindex $i$ stands for radiation, baryons, dark matter or 
dark energy (i.e. the cosmological constant), and the exponent, $I$, takes the following values $I=-4,-3,-3,0$ for the previous species respectively. 
Notice that $\sum_{i=1}^4 \Omega_{i}^{\Lambda CDM} \equiv 1$. The current abundances $\Omega_i^0$ associated with the 
$f(R)$ models are predicted given the initial conditions in the past, since, as we emphasized above, we integrate from the past to the present time. 
Although the initial conditions may vary from model to model, we try to fix them in order to predict the actual abundances today. Nevertheless the 
fixing is not exactly the same for all the models, and so the current abundances will change slightly from model to model. 
We have taken into account the radiation contribution, however, it is almost negligible during the baryon-DM and dark energy domination eras. 
We notice that the Starobinsky and Hu--Sawicky models behave very much like the $\Lambda CDM$ model of GR. In particular, the Hu--Sawicky model is almost indistinguishable 
from the $\Lambda CDM$ model. However, the MJW model shows some differences at higher redshifts. 

The three models exhibit an adequate matter domination era, at least in the range of redshifts explored in the numerical evolution, which is followed by 
an appropriate accelerated expansion afterwards. This is a very important test in view of the results of 
Amendola {\it et al.}~\cite{Amendola2007a,Amendola2007b,Amendola2007c} who showed that several $f(R)$ models are simply unable to recover an adequate matter 
dominated epoch or a suitable accelerated expansion. 
Furthermore, we see that $\Omega_{X}$ decreases like $\Omega_{\Lambda}$ for large $z$, and if we extrapolate this behavior to the very early 
Universe, it is expected that radiation will dominate and hence, that the standard predictions of primordial nucleosynthesis will not be 
spoiled~\footnote{If one does not want to extrapolate any behavior whatsoever, then one needs to integrate the equations from the nucleosynthesis time, where 
radiation dominates, to the current time, by fixing the initial conditions at nucleosynthesis by demanding a successful abundance of the primordial light 
elements and in addition a successful matter and dark-energy domination at late times. This cannot be an easy task, but it may certainly be performed.}.

\bigskip

\begin{figure}[ht]
\includegraphics[width=9cm]{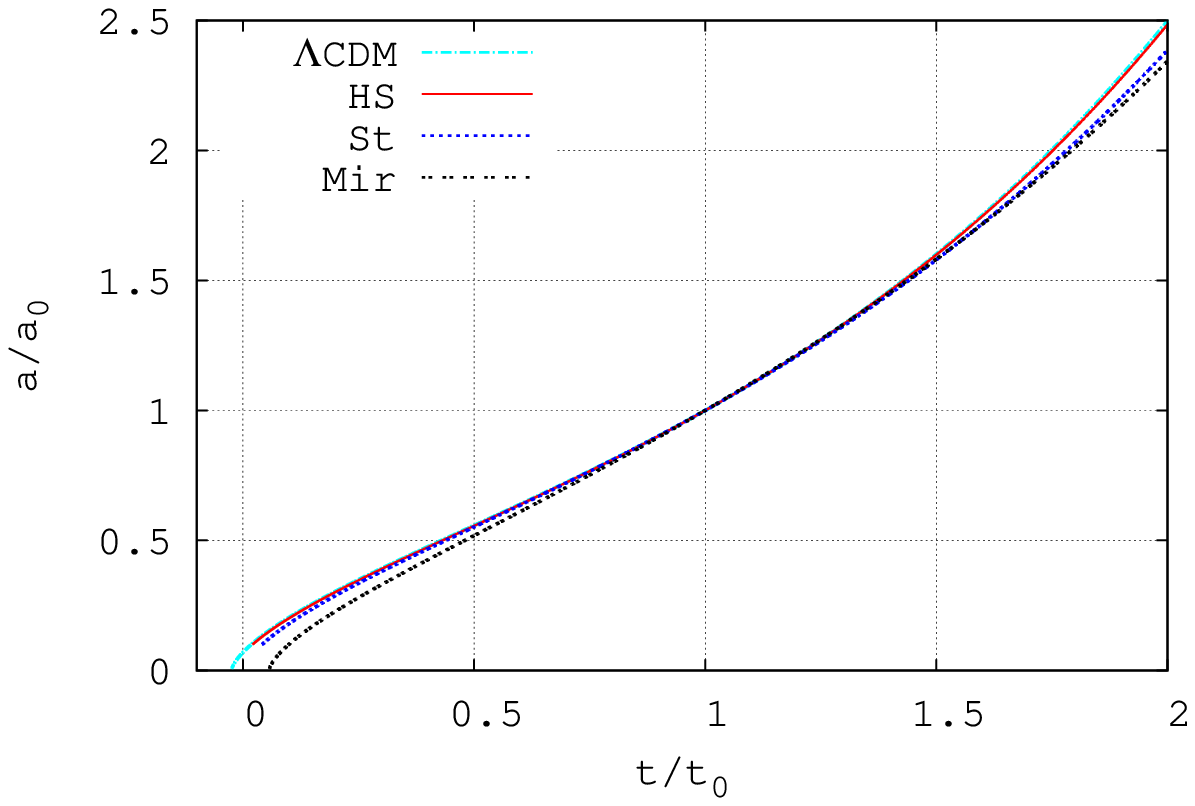}
\caption{Scale factor as a function of cosmic time for the MJW, Starobinsky and Hu--Sawicky models. The $\Lambda CDM$ model is included for reference. 
Here $t_0= H_0^{-1}$. The age of the Universe can be estimated from the time when $a/a_0=1$ (today) and the time when $a/a_0 \sim 0$ (``big bang'').}
\label{fig:avst}
\end{figure}

Figure~\ref{fig:avst} depicts the behavior of the scale factor during the cosmic evolution for the three $f(R)$ models and it is compared with the 
$\Lambda CDM$ model of GR. The curves show the initial epoch at which the equations started to be integrated 
(see the discussion at the end of Sec.~\ref{sec:lumdist}). 
The figure shows that in the Starobinsky and Hu-Sawicky models the age of the Universe is of the order $H_0^{-1}\approx 9.78 h^{-1}\times 10^{9}{\rm y}$ 
(where $h= (H_0/100) {\rm km}^{-1}\,{\rm s}{\rm Mpc}$). While in the case of the MJW model, the Universe is slightly younger, 
$0.94 H_0^{-1}\approx 9.19 h^{-1}\times 10^{9}{\rm y}$. Taking $h=0.7$ as in Figure~\ref{fig:dL1} we obtain an age 
$\sim 13.97 \times 10^{9}{\rm y}$ for the Starobinsky and Hu-Sawicky models and $\sim 13.13 \times 10^{9}{\rm y}$ for 
the MJW model; both agree with estimates of globular clusters in the Milky Way~\cite{Krauss2003}.

\begin{figure}[ht]
\includegraphics[width=9cm]{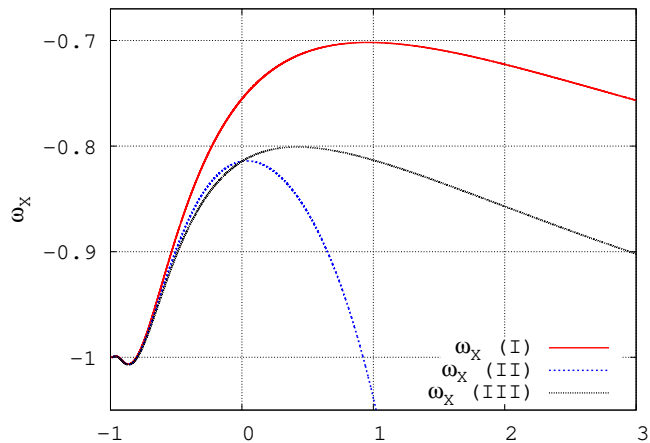}
\caption{Equations of state according to Recipes I--III defined in the main text for the MJW model.}
\label{fig:wxallMir}
\end{figure}

\begin{figure}[ht]
\includegraphics[width=9cm]{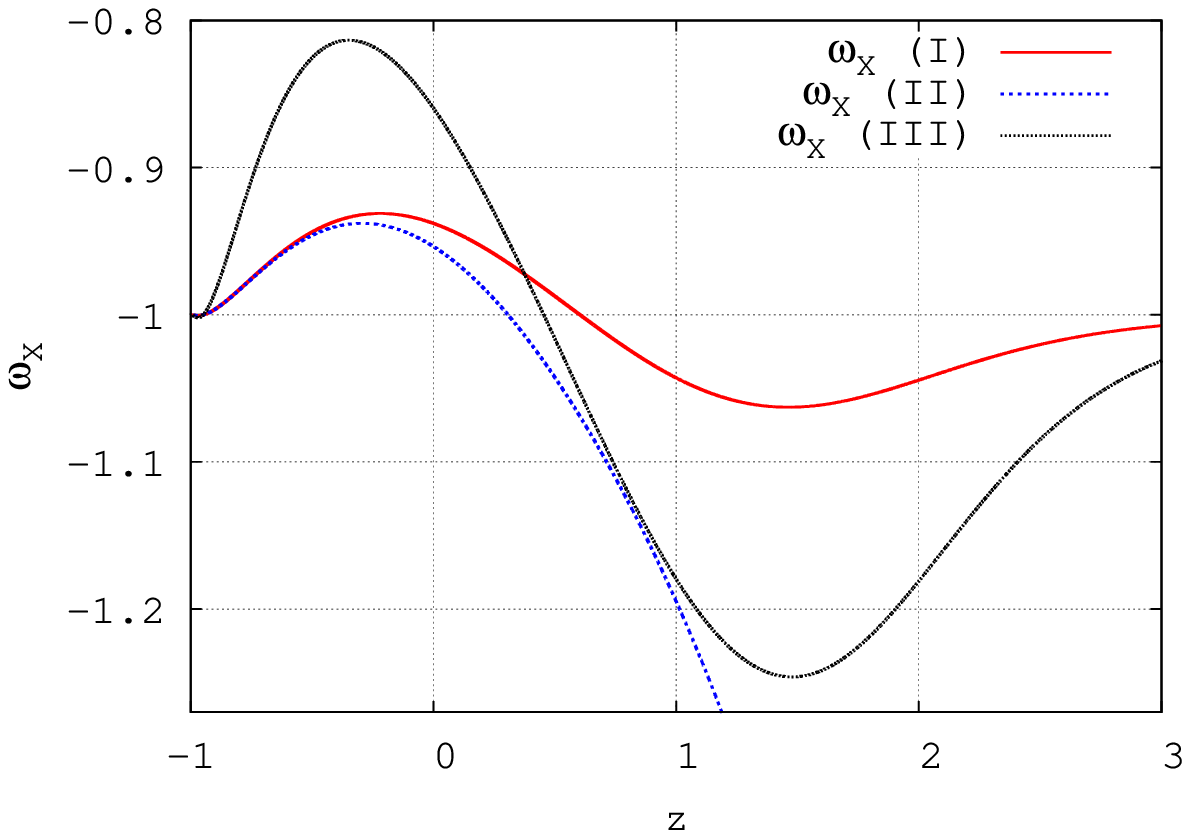}
\caption{Same as Fig.~\ref{fig:wxallMir} for the Starobinsky model.}
\label{fig:wxallSt}
\end{figure}

\begin{figure}[ht]
\includegraphics[width=9cm]{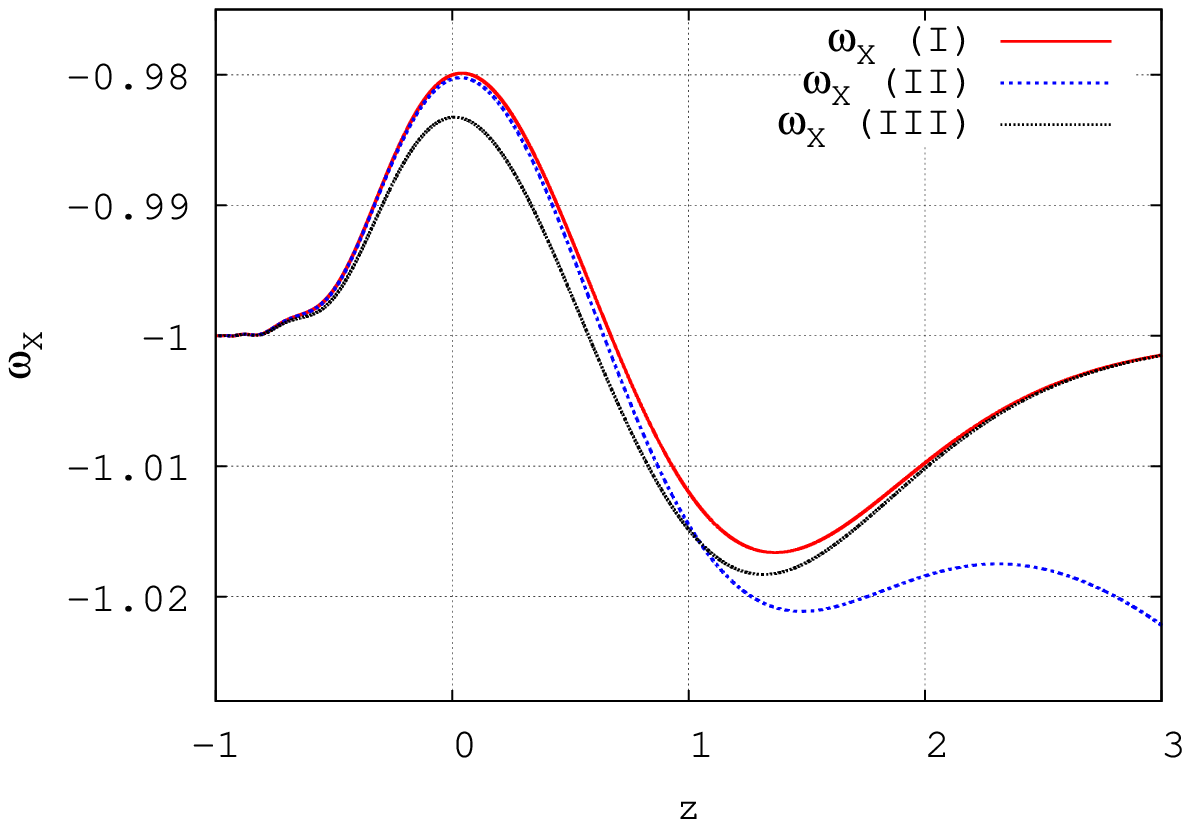}
\caption{Same as Fig.~\ref{fig:wxallMir} for the Hu--Sawicky model.}
\label{fig:wxallHu}
\end{figure}

\begin{figure}[ht]
\includegraphics[width=9cm]{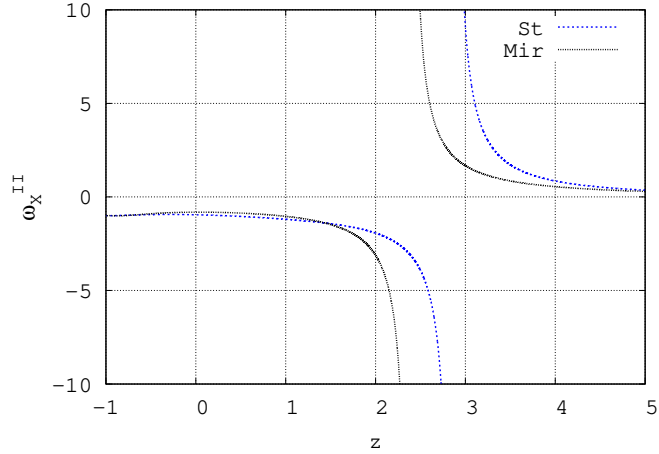}
\caption{Equations of state according to Recipe II defined in the main text for the Starobinsky and MJW models. Notice the divergence due to 
the corresponding energy-density becoming zero (see Figs.~\ref{fig:rhoXMir} and \ref{fig:rhoXSt}). For lower $z$ these EOS are also plotted in 
Figs.~\ref{fig:wxallMir} and \ref{fig:wxallSt}.}
\label{fig:wx2-St-Mir}
\end{figure}

\begin{figure}[ht]
\includegraphics[width=9cm]{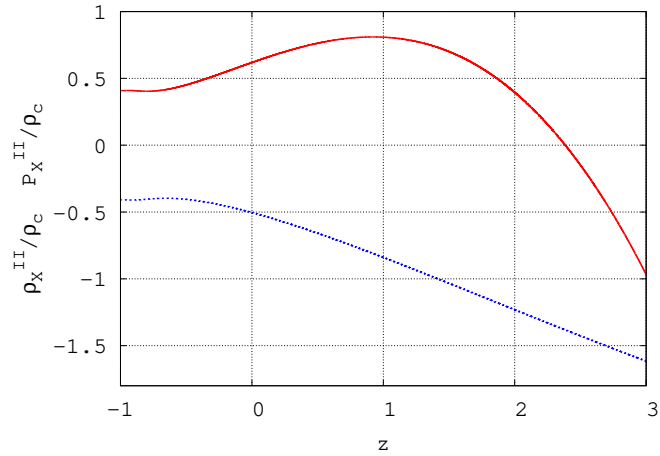}
\caption{Energy-density $\rho_X^{II}$ (solid line) and pressure $p_X^{II}$ (dashed line) computed from the MJW model using Recipe II. 
The quantities are given in units of the today's critical energy-density. Notice that $\rho_X^{II}$ becomes zero at $z\approx 2.38$.}
\label{fig:rhoXMir}
\end{figure}

\begin{figure}[ht]
\includegraphics[width=9cm]{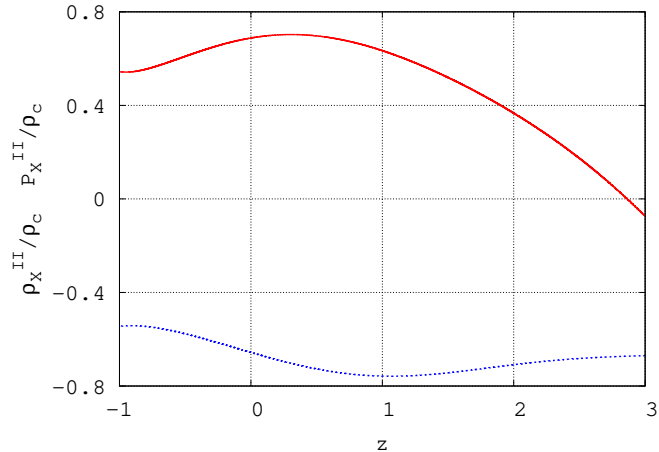}
\caption{Same as Fig.~\ref{fig:rhoXMir} for the Starobinsky model. In this case $\rho_X^{II}$ becomes zero at $z\approx 2.86$.}
\label{fig:rhoXSt}
\end{figure}

\bigskip
Figures \ref{fig:wxallMir}--\ref{fig:wxallHu} show the behavior of the EOS associated with the $X$--fluid based upon the Recipes I--III. 
As concerns the Starobinsky and the Hu--Sawicky models, our 
results are consistent with those reported in Refs.~\cite{Hu2007,Motohashi2010,Motohashi2011a,Motohashi2011b,Bamba} for similar values of the parameters 
where Recipe I was used. We also obtain similar results for the MJW model as compared with Ref.~\cite{Miranda2009}, where Recipe I was used as well. 
The inequivalent definitions of the EOS given by Recipes I and III give similar results for the three models. However, the EOS of Recipe II can 
have a completely different behavior as it diverges at some redshift for the Starobinsky and the MJW models (see Fig.~\ref{fig:wx2-St-Mir}). 
This divergence, which was reported in Ref.~\cite{Amendola2008} and commented in Ref.~\cite{Starobinsky2007}, 
is due to the fact that $\rho_X$ becomes null and then negative, as it is explicitly shown in Figures~\ref{fig:rhoXMir} and \ref{fig:rhoXSt}. 
This in turn can be understood by looking at Eq.~(\ref{rhoX2}), notably at the term $\rho (1-f_R/A)$. First, in the case where $A=1$, 
which corresponds to $\rho_X$, the term $\rho (1-f_R)$ is always positive since $0<f_R<1$ during the cosmic evolution (c.f. Fig.~\ref{fig:fR}). 
If the other contributions are positive, which is expected since $f_{RR}\ll 1$ and since the other terms will give rise to $\Lambda_{\rm eff}>0$, 
then $\rho_X>0$. Now if $A=f_R^0<1$ which corresponds to $\rho_X^{II}$, the term $\rho (1-f_R/f_R^0)$ can be negative in epochs where 
$f_R>f_R^0$ which are precisely those epochs of large $R$ and large $z$. In the matter domination era the negative term $\rho (1-f_R/f_R^0)$ can 
dominate over the geometric $f(R)$-terms of Eq.~(\ref{rhoX2}). This is exactly what happens as depicted in Figures~\ref{fig:rhoXMir} and \ref{fig:rhoXSt}. 
As the evolution continues, $\rho$ decreases to a point where the other terms of Eq.~(\ref{rhoX2}), which increase, balance exactly to give 
$\rho_X^{II}=0$. As the evolution continues to a point where $f_R<f_R^0$ or $\rho$ is sufficiently small, the term $\rho (1-f_R/f_R^0)$ becomes positive or 
small, and then $\rho_X^{II}$ becomes positive. All this behavior exacerbates as the parameter $f_R^0$ differs sufficiently from unity, since then the 
term $\rho (1-f_R/f_R^0)$ becomes important. 
Since in the Starobinsky and MJW models, unlike the Hu--Sawicky one, the parameter $f_R^0$ is not fixed in advance but rather predicted from the 
initial conditions in the past, then for those models it turns that $f_R^0$ differs from unity more importantly than in the Hu--Sawicky case where 
the model was constructed in such a way that at present time $f_R^0\approx 1$ (see Table~\ref{tab:todayparam}). In the Hu--Sawicky model then the term $\rho (1-f_R/f_R^0)$ 
is not very important in the epochs where $\rho$ is not very large since the factor $1-f_R/f_R^0\approx 0$. However, farther in the past 
where $\rho$ dominates, the term $\rho (1-f_R/f_R^0)$ might become important and can make $\rho_X^{II}=0$ in that model too. This depends on the rates at which the factor 
$1-f_R/f_R^0$ and the energy-density $\rho$ decreases and increases respectively as looking from the present to the past. In any instance, 
this behavior is unacceptable and indicates that the definition of EOS II is not adequate.

Like in Refs.~\cite{Hu2007,Martinelli2009,Motohashi2010,Motohashi2011a,Motohashi2011b,Bamba}, we also find oscillations of the EOS around the 
``phantom divide'' $\omega_X= -1$, for the Starobinsky and Hu--Sawicky models. However, for the MJW model, the well behaved EOS, $\omega_X$ 
and $\omega_X^{III}$ do not cross the phantom divide 
in the past. This behavior was already remarked in \cite{Miranda2009}.

\begin{figure}[ht]
\includegraphics[width=9cm]{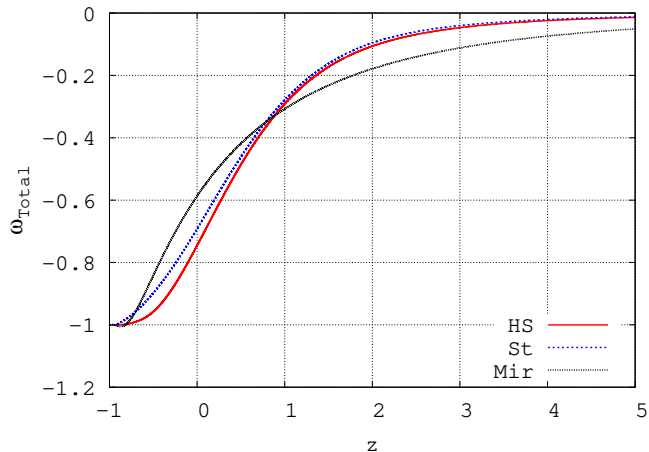}
\caption{Equation of state $\omega_{\rm tot}$ as defined in Eq.~(\ref{wtot}).}
\label{fig:wtot}
\end{figure}

\begin{figure}[ht]
\includegraphics[width=9cm]{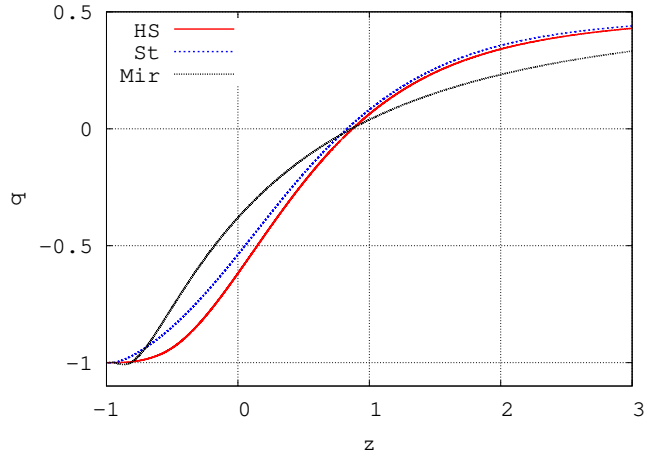}
\caption{The deceleration parameter $q$ for the  MJW, Starobinsky and Hu--Sawicky models.}
\label{fig:q}
\end{figure}

\begin{figure}[ht]
\includegraphics[width=9cm]{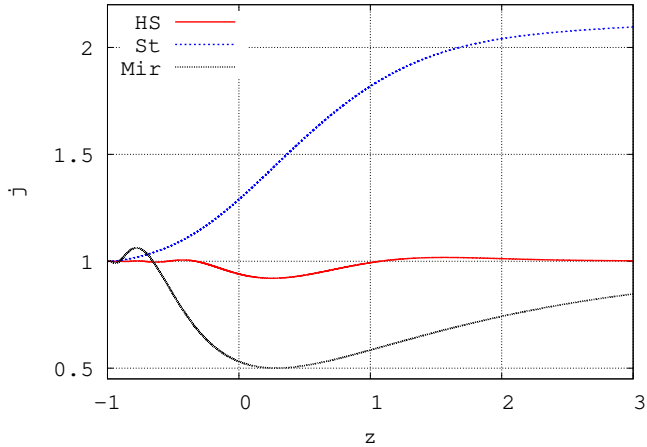}
\caption{The jerk parameter $j$ for the  MJW, Starobinsky and Hu--Sawicky models.}
\label{fig:j}
\end{figure}

Figure~\ref{fig:wtot} depicts the total EOS (i.e. using Recipe I). We appreciate the transition from the matter dominated era to the 
accelerated era. Figures~\ref{fig:q} and \ref{fig:j} depict the deceleration parameter and the jerk as computed from Eqs.~(\ref{dec}) and (\ref{jerk}) for the 
three $f(R)$ models. Table~\ref{tab:todayparam} shows the current values of these and other quantities for such models.

\begin{table}
\centering
\begin{tabular}{|c| |c| |c| |c| |c| }
\hline
$q_{0}$ & $j_{0}$ & $\frac{R_{0}}{H_0^2}$ & $f_R^{0}$ & model \\
\hline
-0.618 & 0.94 & 9.74 & 0.99 &  Hu--Sawicky  \\
-0.54 & 1.29 & 2.20 & 0.96  &  Starobinsky \\
-0.38 & 0.53 & 8.91 & 0.80  &  MJW \\
\hline
\end{tabular}
\caption{Deceleration, jerk and other quantities of the models at $z=0$ (today).}
\label{tab:todayparam}
\end{table}

\subsection{Luminosity distance and SNIa}
\label{sec:lumdist}

The accelerated expansion of the Universe was corroborated by the measurements of luminous distance in supernovae (SNIa), when comparing with the 
predictions of the $\Lambda CDM$ of GR. In which follows we confront the same data with the accelerated expansion predicted by $f(R)$ theory. 
We compute first the luminous distance given by
\begin{equation}
\label{DL}
d_{L}^{\rm flat}=  \frac{\zeta({\bar a})}{\bar a}\,\,,
\end{equation}
where ${\bar a}= a/a_0$ and 
\begin{equation}
\label{chi}
\zeta= c\,H_0^{-1} \int^{1}_{\bar a}\frac{d{\bar a}^*}{ {\bar a}^{*\,2} {\bar H}({\bar a}^*) }\,\,\,,
\end{equation}
where we have introduced explicitly the speed of light $c$ in order to compute the distances in units of ${\rm Mpc}$, and ${\bar H}:= H/H_0$. 
Another useful quantity in cosmology is the angular diameter distance given by~
\begin{equation}
\label{DA}
D_{A}^{\rm flat}= {\bar a} \zeta({\bar a}) \,\,.
\end{equation}
We emphasize that the above expressions for the luminous and  angular diameter distances are valid only for $k=0$.

Perhaps the easiest way to compute $\zeta$ within the framework of the numerical scheme that we have devised is to transform Eq.~(\ref{chi}) into 
the following differential equation for $\zeta$:
\begin{equation}
\frac{d {\bar \zeta} }{d{\bar a}}=-\frac{1}{{\bar a}^2{\bar H}({\bar a})}\,\,\,,
\end{equation}
where ${\bar \zeta}= \zeta/(c\,\,H_0^{-1})$ is dimensionless and which in terms of the variable $\alpha= {\rm ln}({\bar a})$ defined in Eq.~(\ref{alpha}), reads
\begin{equation}
\label{chidif}
{\bar \zeta}' =-\frac{e^{-\alpha}}{{\bar H}}\,\,\,.
\end{equation}
This first order differential equation is integrated simultaneously with the field equations in the way we just described.  
A technical but important point is that {\it a priori} we do not know what the initial value of ${\bar \zeta}$ is at the past epoch where the numerical integration 
starts. However, that value is easily found by a shooting method such that ${\bar \zeta}$ is zero today (at $z=0=\alpha$). Given ${\bar \zeta}$ one then computes 
$d_{L}^{\rm flat}$ and $D_{A}^{\rm flat}$ from Eqs.~(\ref{DL}) and ~(\ref{DA}). Actually the quantity which is usually reported is not $d_{L}^{\rm flat}$ but the 
{\it distance modulus} given by
\begin{equation}
\mu:= m-M= 5{\rm log}_{10} (d_{L}^{\rm flat}/{\rm Mpc}) + 25 \,\,\,.
\end{equation}

\begin{figure}
\includegraphics[width=9cm]{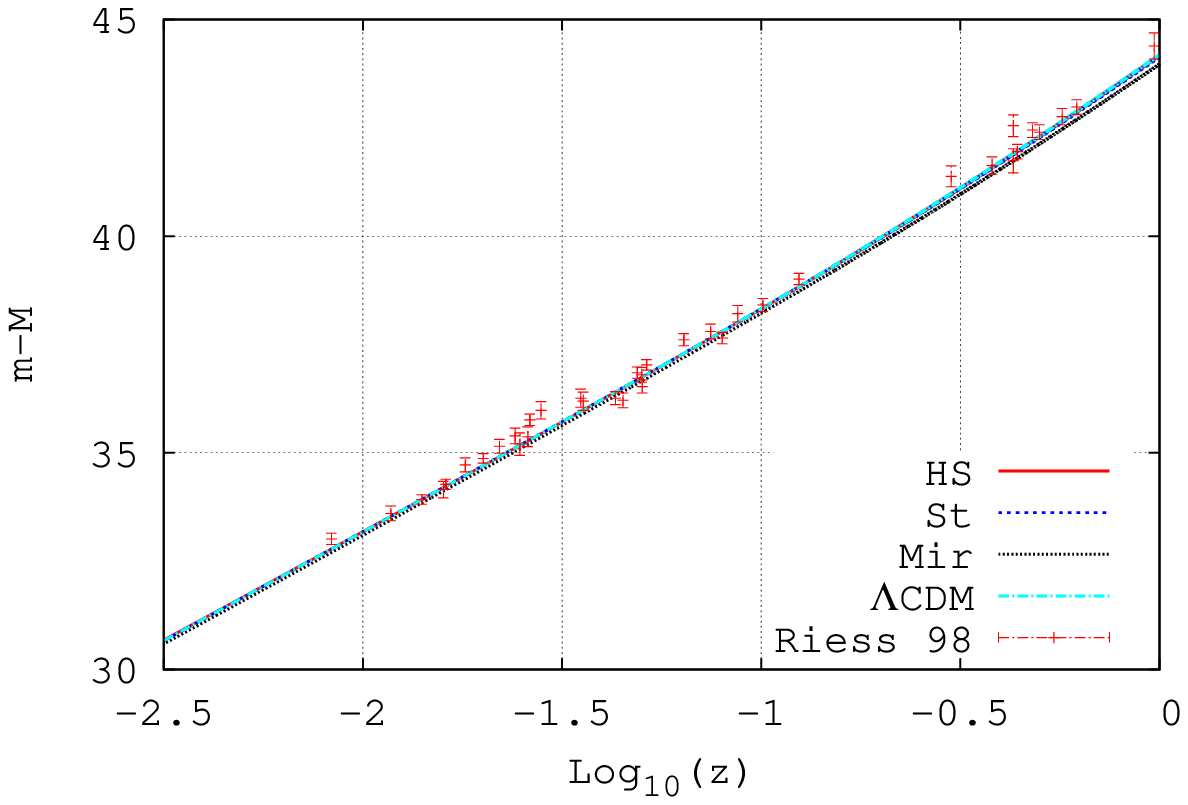}
\caption{Distance modulus for the MJW, Starobinsky and Hu--Sawicky models. The $\Lambda CDM$ model is also plotted for reference. The data was taken from 
Riess {\it et al.}~\cite{Riess1998}. Here $H_0= 70\,{\rm km}\,{\rm s}^{-1}{\rm Mpc}^{-1}$.}
\label{fig:dL1}
\end{figure}

\begin{figure}
\includegraphics[width=8.5cm]{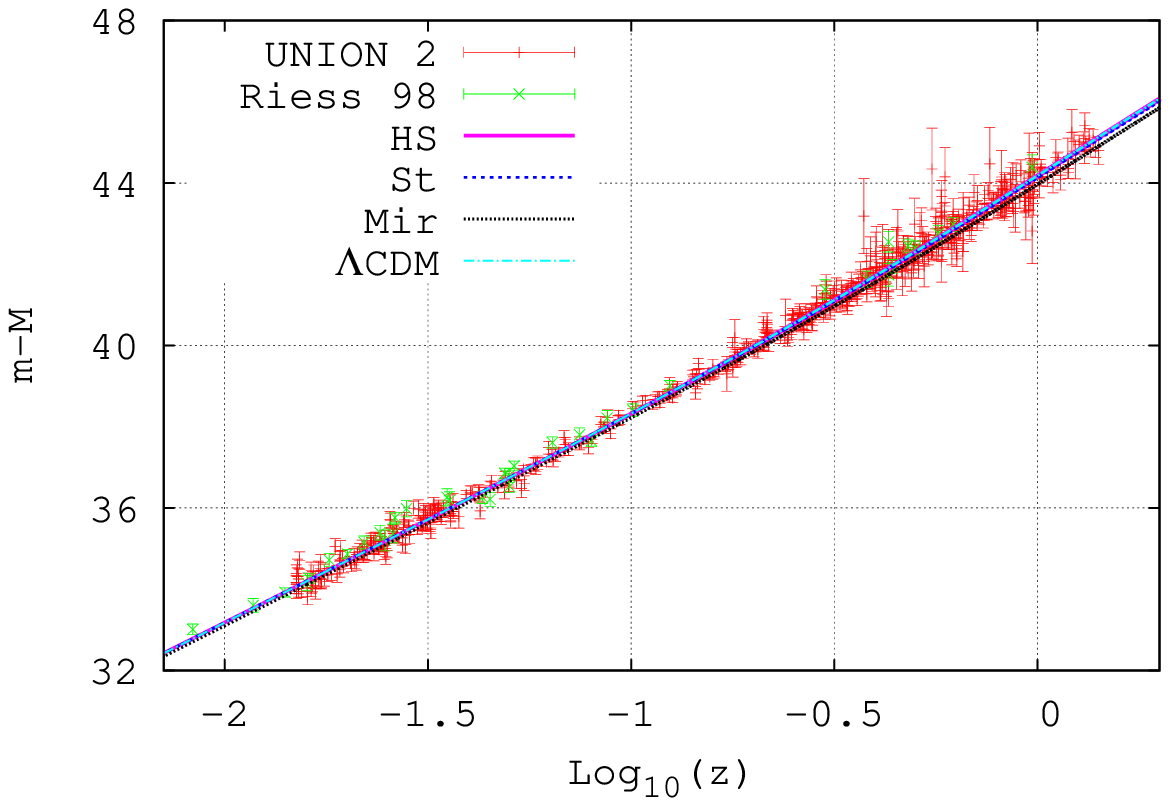}
\includegraphics[width=8.5cm]{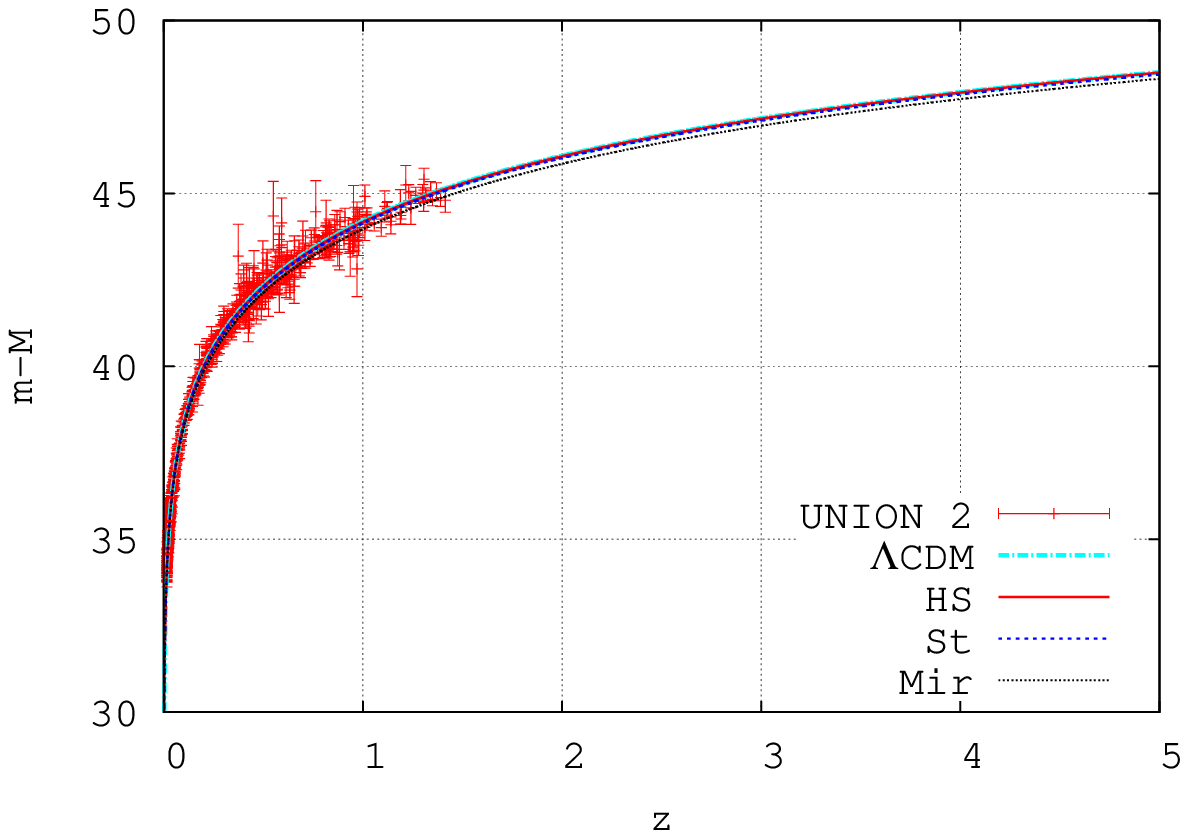}
\caption{Left: Same as Fig.~\ref{fig:dL1} including the Riess {\it et al.} data (green)~\cite{Riess1998} and the Union 2 compilation of Amanullah {\it et al.} 
(red)~\cite{Amanullah2010} (color online). Right: similar as the left box but using a different scale and larger $z$ (only Union 2 data is considered here).}
\label{fig:dL1Union}
\end{figure}

\begin{figure}
\includegraphics[width=9cm]{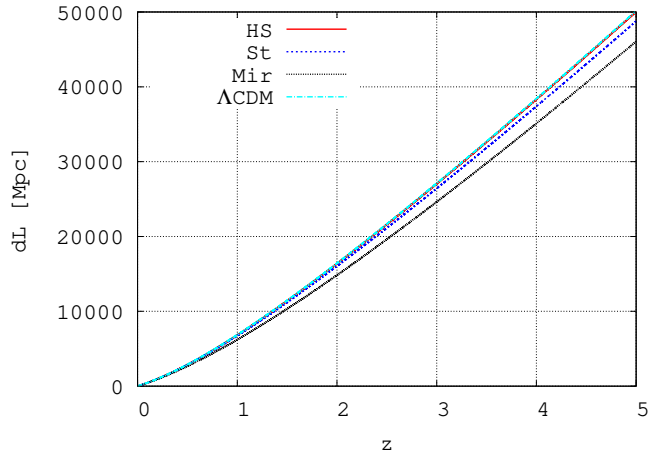}
\caption{Similar to Fig.~\ref{fig:dL1Union} for the luminous distance and larger $z$ (observational data is not included here).}
\label{fig:dL2}
\end{figure}

\begin{figure}
\includegraphics[width=9cm]{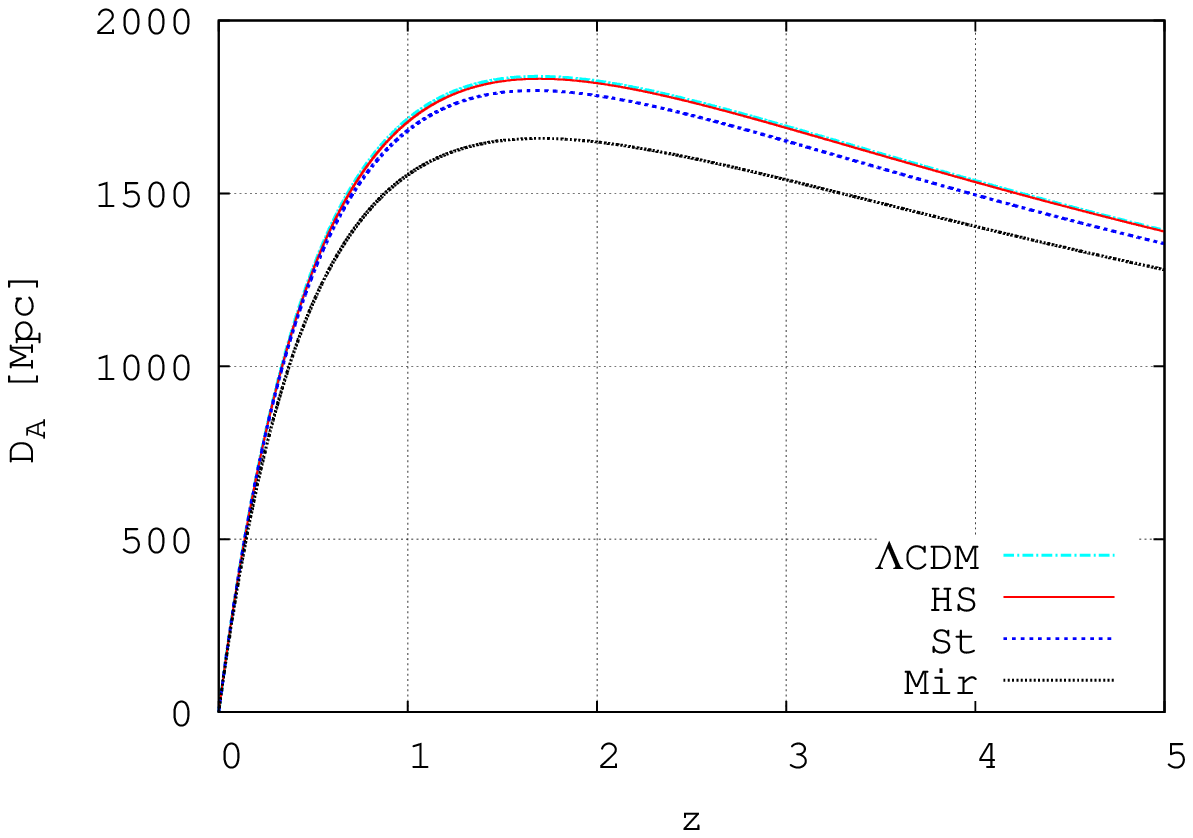}
\caption{Angular diameter distance for the MJW, Starobinsky and Hu--Sawicky models. The $\Lambda CDM$ model is also plotted for reference.}
\label{fig:DA}
\end{figure}

Figures~\ref{fig:dL1} and \ref{fig:dL1Union} show the luminosity distance for the three $f(R)$ models described above and are compared with the $\Lambda CDM$ model. 
Here the data corresponds to both the ``historical'' data of Riess {\it et al.} of Ref.~\cite{Riess1998} and also the UNION 2 data of Amanullah 
{\it et al.}~\cite{Amanullah2010}. We appreciate that up to $z=1$ there is no significant difference between the three $f(R)$ models 
and the $\Lambda CDM$ model. Nonetheless, for larger $z$ (see Fig.~\ref{fig:dL2}), significant differences arise which in the future could constrain or rule out these or 
other $f(R)$ models or even challenge the $\Lambda CDM$ paradigm. Figure~\ref{fig:DA} depicts the angular diameter distance.

\begin{figure}[ht]
\includegraphics[width=9cm]{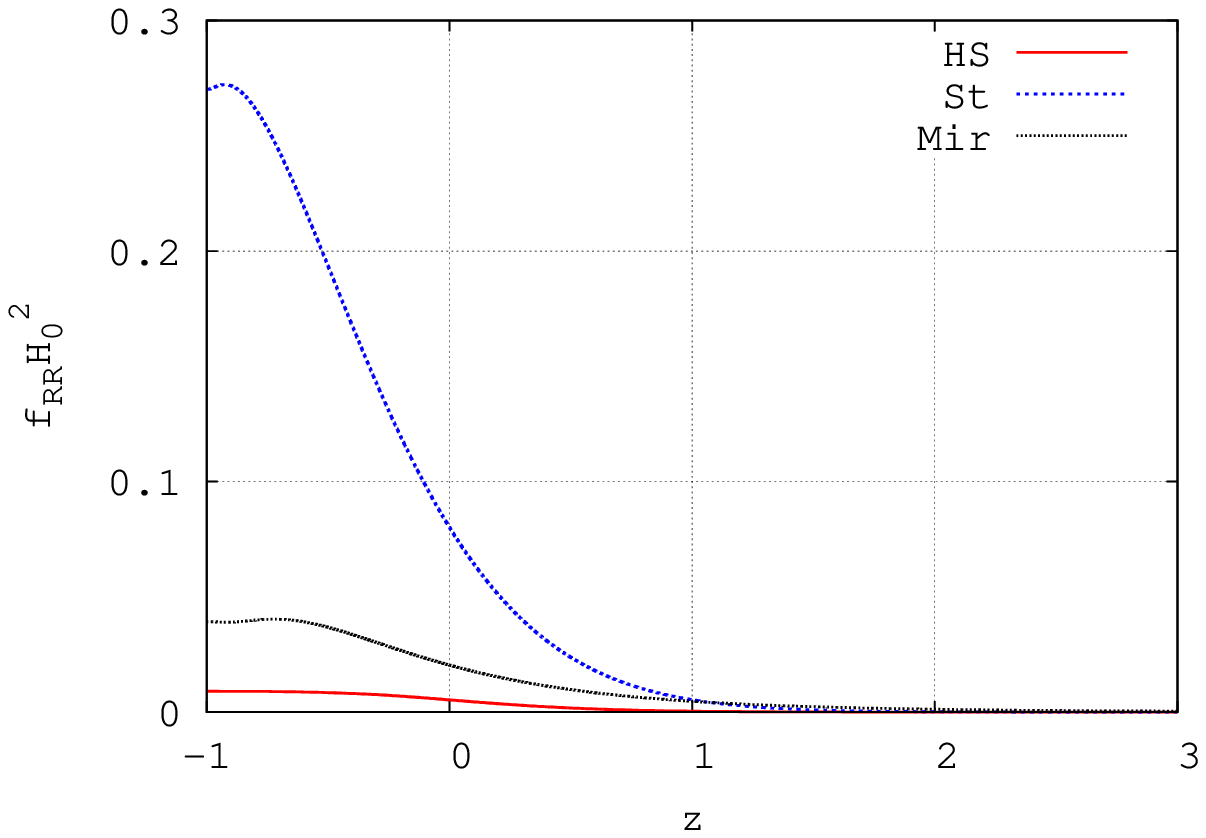}
\caption{The second derivative $f_{RR}:= d^2f/dR^2$ of Fig.~\ref{fig:fRR} as a function of $z$. Notice that $f_{RR}$ goes to zero for large $z$.}
\label{fig:f2vsz}
\end{figure}

\begin{figure}[ht]
\includegraphics[width=9cm]{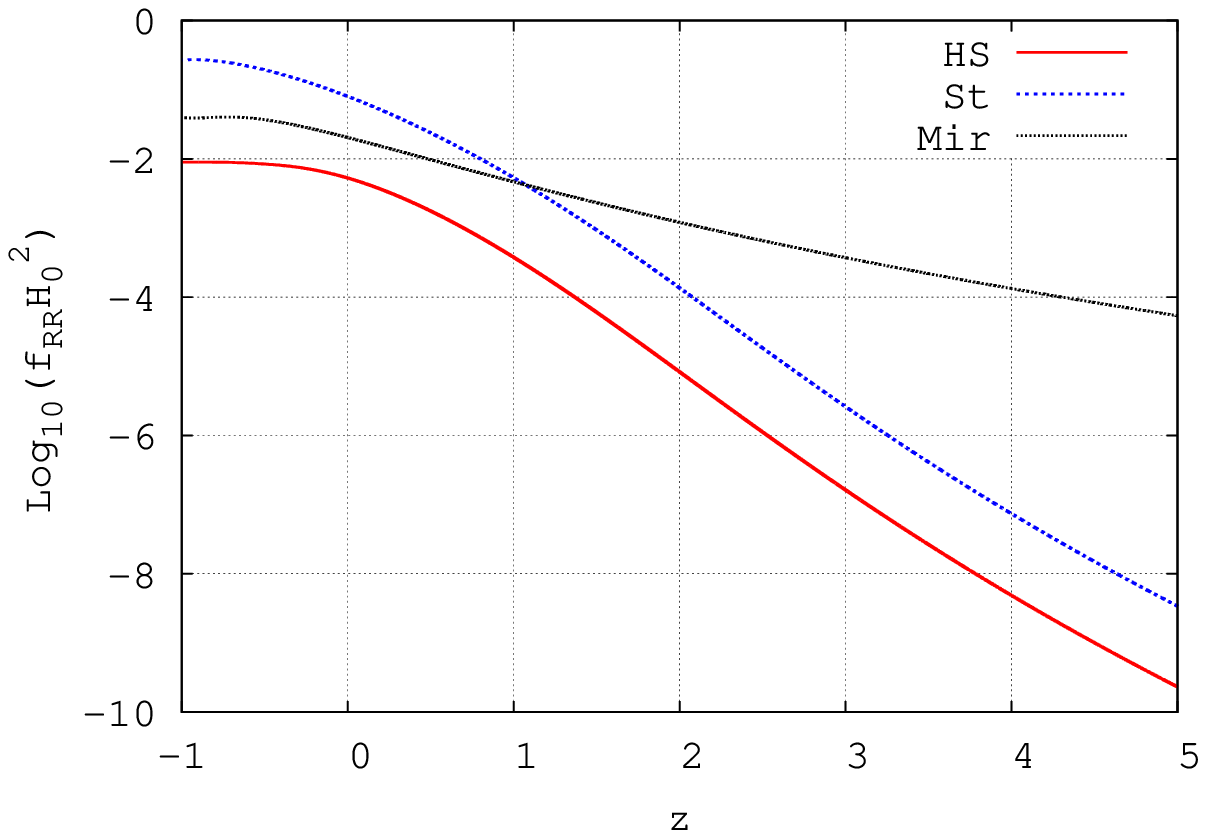}
\caption{Same as Fig~\ref{fig:f2vsz} in logarithmic scale.}
\label{fig:logf2vsz}
\end{figure}

To conclude this section we mention a technical point that in principle restricts the integration of the equations for some $f(R)$ models and which manifest 
specifically in the Starobinsky and Hu--Sawicky case where $f_{RR}$ approaches zero for large $R/H_0^2$. This regime occurs during the cosmic evolution for 
large $z$. Figures~\ref{fig:f2vsz} and \ref{fig:logf2vsz} depict the behavior of $f_{RR}$, where we appreciate that $f_{RR} H_0^2\ll 1$, and approaches zero 
much faster in the Starobinsky and Hu--Sawicky models than in the MJW model. This behavior is consistent with Figures~\ref{fig:F(R)'s} and \ref{fig:fRR} where 
one sees that for large $R$ the Starobinsky and Hu--Sawicky models behave as $f(R)\approx R -2 \Lambda_{\rm eff}$, and therefore $f_{RR}\approx 0$. 
Now, the point is that when $f_{RR}\approx 0$ in the past, Eq.~(\ref{R-numerical}) develops large variations as $f_{RR}^{-1}$ appears in 
the r.h.s and affects the numerical precision. Since in the MJW model $f_{RR}$ decreases slower than in the other two models, we have corroborated that for 
the former we can perform the integration starting quite far in the past (even at the recombination epoch) without finding any numerical troubles. 
Actually this feature can also be appreciated in Figure~3 of Ref.~\cite{Miranda2009}, where the integration starts even before recombination.

\section{Discussion}
\label{sec:discussion}

Modified theories of gravity, like metric $f(R)$ theories, have been analyzed thoroughly in recent years mainly to explain the accelerated expansion 
of the Universe which was inferred from the measurements of luminous distance of SNIa, while some other theories have been put 
forward to avoid the need of dark matter. In this article we have focused mainly on the mechanism to produce an accelerated expansion. 
Such behavior can be most easily accounted 
within the general theory of relativity and by (re)introducing the cosmological constant which was proposed 
by Einstein almost one hundred years ago. The motivation behind the proposal for modifying GR was to avoid the introduction of such constant and then 
to circumvent the apparent problems associated with it. Furthermore, such an alternative exempts us from adding new fields (scalar or 
otherwise) like quintessence or k-essence, in order to explain that acceleration. Nevertheless this alternative adds, to our opinion 
much more troubles than solutions. While many specific $f(R)$ models are able to produce an accelerated expansion similar to the $\Lambda CDM$ model, 
they have, at the same time, spoiled many of the successes of GR or are inconsistent with other features of cosmology, like an adequate matter 
dominated era. Only a few models have succeeded in explaining, at least partially, the actual 
cosmological evolution of the Universe without disturbing, for instance, the predictions at Solar System scales. Since, $f(R)$ theories do not 
introduce a new fundamental principle of nature, there is then, not a deep criteria that favors one among the apparently viable $f(R)$ models. 
Basically some kind of ``handcraft''  have been used so far to mold specific $f(R)$ or in other cases even reconstruction methods~\cite{Dunsby2010,reconstruction}. 
At any rate, simplicity would be in favor of GR with $\Lambda$. 
In this article we argued that even if this kind of modified theories can be internally consistent, some special care has to be taken into 
account when they are analyzed. For instance, it was ``discovered'' that $f(R)$ theories can be equivalent to scalar-tensor theories of gravity. This 
identification is free of inconsistencies provided that the mapping between both representations is well defined. For that it is required that the 
function $f_R$ be a monotonic function of the Ricci scalar $R$. 
Several of the apparently successful models fail to fulfill this condition in general, 
and yet, different authors have used the STT representation. One of the consequences is that the scalar field potential turns to be multivalued. This 
pathology has contributed to create confusion in the subject, in addition to the already existing confusion between frames in STT. We have emphasized 
that in order to avoid such potential drawbacks $f(R)$ theories should be treated using the original variables. This is not only possible, but 
in our view, it turns to be much more transparent, even if the equations seem more involved at first sight. Although some other people share 
this view and have used the original variables in several applications, we give a step forward and propose here 
a general system of equations simpler than the one usually used, and when applied particularly to cosmology, it reduces further and can be treated numerically 
as an initial value problem, where the initial values are restricted by a modified Hamiltonian constraint. Previously we presented a system of equations that can be 
used to construct compact objects in static and spherically symmetric spacetimes and showed the way to treat them numerically~\cite{Jaime2011}. 
In the current article we analyzed three specific $f(R)$ models that are viable, at least in the background, since they 
provide an adequate matter dominated behavior followed by a correct accelerated expansion. It has been argued that 
the Miranda {\it et al.} model~\cite{Miranda2009} is ultimately incompatible when cosmology is analyzed at the level of perturbations or in 
the Solar System~\cite{delaCruz2009}, but very likely a deeper analysis is required in order to rule out this model completely~\cite{Miranda2009b}. 
In the future we plan to analyze some other models, like the exponential ones~\cite{Zhang2006,exponential,gravwaves1} which can be viable as well. 
For each of the three classes of $f(R)$ models that we considered here, we analyzed three possible inequivalent equations of state associated with the modified 
gravity and which have been studied in the past by several authors. Such EOS arise from  general 
EMT that represent the modified gravity or geometric dark energy, although such interpretation is to be handled with care 
as several authors have warned before. We have identified at least four inequivalent definitions of the EMT which 
lead to the three EOS just alluded. Two of such EMT's have the unpleasant feature that are not conserved. One of the other four EOS 
can diverge at some red-shift. Thus the more appealing definition comes from Recipe I whose EMT is conserved and the corresponding EOS behaves 
appropriately. We emphasized that it is important to come to an agreement on 
which definition of the EOS will be ultimately compared with the cosmological observations. This is crucial in view that the forthcoming 
data might differentiate between them as precision is increased~\cite{BB,obsEOS}. 

It is natural to ask about the new predictions that $f(R)$ 
theories can make in addition to explaining the accelerated expansion while reproducing, in the viable cases, the previous successes of GR and 
in particular the $\Lambda CDM$ model. There are indeed several predictions that, in principle, would allow us to distinguish between GR and $f(R)$ 
theories. One of them is the additional polarization modes of gravitational waves like the  
``breathing'' mode~\cite{gravwaves1,gravwaves2}. At present time the current gravitational-wave detectors are not sensitive enough 
to detect gravitational radiation and therefore, this feature cannot be used as a tool to falsify alternative theories 
of gravity yet, but in a near future this will be certainly possible~\cite{Will}. The binary pulsar on the other 
hand can be an excellent tool to do so at present time. We know that the binary pulsar can constrain STT even if they succeed in passing the 
Solar System tests~\cite{Damour96}, thus, the same might happen for $f(R)$ theories. 
At the cosmological level, $f(R)$ theories predict a large-scale integrated Sachs--Wolf effect which is different from the one predicted in GR~
\cite{Zhang2006,Zhao2010,ISW,Bertschinger2008}. 
This is because the total EMT that one can associate with such theories is not necessarily ``isotropic'' at the time of recombination and thus 
the metric potentials of scalar perturbations in the Newtonian gauge are not equal (in absolute value). Nevertheless, due to the 
so called cosmic variance, it turns difficult to use it as a further constraint for some values of the model parameters. On the other hand, 
weak and strong gravitational lensing as well as the growth of matter perturbations can also be affected when those potentials are different
~\cite{Zhang2007,Zhao2010,Bertschinger2008,HS&Staromodel}, and so they can further constrain the specific models.

In this article we have not attempted to best-fitting the parameters using current observations, but rather using the $\Lambda CDM$ models as a 
standard, taking a zero spatial curvature as a prior. The former analysis will be pursued in a future work. 


\newpage
\section{Appendix}
\label{sec:appendixA}
\section*{Dimensionless form of the cosmological equations and tests.}
The differential equations (\ref{R-numerical})--(\ref{accmod-numerical}) can be recasted in 
dimensionless form which is more suitable for numerical integration. One can introduce the following dimensionless quantities:
\begin{eqnarray}
{\bar a} &=& a/a_0 \,\,\,,\\
{\bar t} &=& t/H_0^{-1} \,\,\,,\\
{\bar H} &=& H/H_0 \,\,\,,\\
{\bar f} &=& f/H_0^2 \,\,\,,\\
{\bar R} &=& R/H_0^2 \,\,\,,\\
{\bar f_{RR}} &=& f_{RR}/H_0^{-2} \,\,\,,\\
{\bar f_{RRR}} &=& f_{RRR}/H_0^{-4} \,\,\,,\\
{\bar \rho} &=& \rho/\rho_{\rm crit}^0\,\,\,,\\
{\bar p_{\rm rad}} &=& p_{\rm rad}/\rho_{\rm crit}^0\,\,\,,\\
{\bar T} &=& T/\rho_{\rm crit}^0\,\,\,,\\
\end{eqnarray}
where $\rho_{\rm crit}^0:= 3 H_0^2/(8\pi G)$. Notice that $f_R$ is already dimensionless. 
 
In terms of these quantities Eqs.~(\ref{R-numerical})--(\ref{invHubble}) read 
\begin{equation}
\label{R-numericaldim}
{\bar R}''=-{\bar R}' \left(1+\frac{{\bar R}}{6{\bar H}^{2}}\right)-\frac{1}{3{\bar f_{RR}} {\bar H}^{2}}\left[ 
3{\bar f}_{RRR}{\bar H}^{2} {\bar R}'^{2}+ 2 {\bar f}-{\bar f}_R {\bar R}  + 3 {\bar T} \right]\,\,,
\end{equation}

\begin{equation}
\label{dHnumericaldim}
{\bar H}'=-2{\bar H}+\frac{{\bar R}}{6{\bar H}}\,\,,
\end{equation}

\begin{equation}
\label{Friedmod-numericaldim}
{\bar H}^2 + \frac{1}{{\bar f}_R}\left[ {\bar f}_{RR} {\bar H}^{2} {\bar R}' -\frac{1}{6}\left( {\bar f}_R {\bar R}- {\bar f}\right)\right]= 
\frac{{\bar \rho}}{{\bar f}_R}\,\,,
\end{equation}

\begin{equation}
\label{accmod-numericaldim}
{\bar H}' = -{\bar H} + \frac{1}{{\bar f}_{R} {\bar H}}\left( {\bar f}_{RR}  {\bar H}^{2} {\bar R}' + \frac{{\bar f}}{6} - {\bar \rho} \right) \,\,\,\,,
\end{equation}

\begin{eqnarray}
\label{invHubbledim}
\frac{d{\bar t}}{d\alpha}= 1/{\bar H}\,\,\,,
\end{eqnarray}

where we took $T^t_{\,\,t}= -\rho$ and $T= -(\rho_{\rm bar}+ \rho_{\rm DM})$. Notice now, as compared with Eqs.~(\ref{R-numerical}), (\ref{Friedmod-numerical}), 
and (\ref{accmod-numerical}), that the factor $\kappa=8\pi G$ no longer appears, and a factor of `3' appears in Eq.~(\ref{R-numericaldim}) next to ${\bar T}$ 
in order for $T$ to be given in units of $\rho_{\rm crit}^0$. Similarly in Eqs.~(\ref{Friedmod-numericaldim}) and (\ref{accmod-numericaldim}) the factors 
$\kappa/3$ have now disappeared for $\rho$ to be given in units of $\rho_{\rm crit}^0$.
 
\begin{figure}[ht]
 \includegraphics[width=8cm]{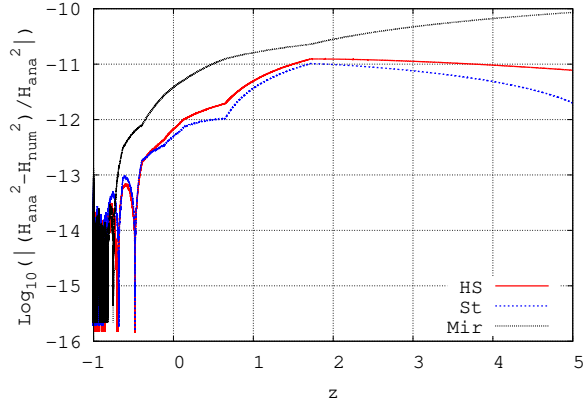}
 \vspace{0.1cm}
 \caption{Relative error commited in the verification of the modified Hamiltonian constraint during the numerical integration for the MJW, Starobinsky and 
Hu--Sawicky models.}
\label{fig:errorrel2}
\end{figure}

As concerns the {\it internal} test performed to check the numerical accuracy and consistency of our computer code, we checked up to which extent the 
modified Hamiltonian constraint Eq.~(\ref{Friedmod-numericaldim}) is verified. We compare the value of ${\bar H}$ given by Eq.~(\ref{Friedmod-numericaldim}) 
and call it ${\bar H}_{\rm ana}$ with the value that arises from integrating Eq.~(\ref{dHnumericaldim}) or (\ref{accmod-numericaldim}), that we call it 
${\bar H}_{\rm num}$. The relative difference between both measures the degree of verification (or failure) of the constraint. Figure~\ref{fig:errorrel2} depicts 
this difference as a function of $z$. Notice that the maximum relative error is lower than $10^{-10}$. At the initial $z$ Eq.~(\ref{Friedmod-numericaldim}) 
is satisfied exactly by construction and thus the precision is infinite. This value is thus not depicted.

We conclude then that results obtained from the FORTRAN code used to solve the differential equations Eqs.~(\ref{R-numericaldim})--(\ref{invHubbledim}) and 
Eq.~(\ref{chidif}) are reliable.


\acknowledgments
This work was supported in part by DGAPA-UNAM grants IN117012-3, IN115310, IN112210, IN110711 
and SEP-CONACYT 132132. L.G.J. acknowledges support from scholarship CEP-UNAM.



\end{document}